\documentclass{PoS}

\title{LHC phenomenology at next-to-leading order QCD: theoretical progress and new results}

\ShortTitle{LHC phenomenology at NLO QCD}
\author{Thomas Binoth\thanks{This work was supported by the British Science and Technology 
Facilities Council (STFC), the Scottish Universities Physics Alliance (SUPA) and
the European {\tt HEPTOOLS} Research Training Network.}\\
        School of Physics and Astronomy\\
        The University of Edinburgh\\
        James Clerk Maxwell Building\\
        The King's Buildings\\
        Mayfield Road\\
        Edinburgh EH9 3JZ\\
        Scotland, UK\\
        E-mail: \email{thomas.binoth@ed.ac.uk}}

\abstract{
In this talk I report on recent developments and results
relevant for LHC phenomenology at next-to-leading 
order QCD. Feynman diagrammatic and unitarity based methods 
have both seen considerable improvements and new ideas recently.  
Current approaches point towards automated evaluation
of one-loop multi-particle amplitudes. Many results for notoriously
difficult processes are under construction by 
several groups and,  given the enormous recent progress, 
it can be concluded that LHC phenomenology 
at full next-to-leading order level will become the standard
approximation soon.}

\FullConference{XII Advanced Computing and Analysis Techniques in Physics Research\\
		 November 3-7 2008\\
		 Erice, Italy}
\begin{document}%

\section{Introduction}%

With the start of the Large Hadron Collider (LHC) at CERN we enter exiting times 
in particle physics. Once the machine is fully operating we will 
be able to explore an energy range which has never been reached 
in a laboratory. Many people in this audience are involved in the 
enormous challenge to master the huge amount of data which will be
produced during collisions. I will focus in this talk on 
the theoretical challenge of predicting LHC phenomena as
precisely as possible using our good old Standard Model. 
We all very much hope for a discrepancy between well-understood data sets
and sufficiently precise theoretical predictions.    
Such a discrepancy will be a signal of new and maybe completely 
unexpected  phenomena which will help us to push our understanding 
of nature. 

What should be  expected from the LHC?
The first and most important issue is
 the test of the Higgs mechanism within the Standard Model
 which predicts the prominent Higgs boson. The LEP experiments set a lower bound of 114.4 GeV 
on its mass from 
direct searches. 
As a Higgs boson with a mass higher than about 200 GeV would 
induce radiative corrections which are in conflict with 
electroweak precision measurements, the Standard Model Higgs
boson is pretty much nailed down to a low mass window. 
Ironically, this window is not as easy to close as the mass range above
the two Z boson threshold of about 180 GeV, where two muon pairs would 
provide a gold-plated discovery mode for the Higgs boson.  
The identification of  any signal with the Standard Model Higgs boson
will also entail detailed measurements of quantum numbers like
spin and CP properties \cite{Djouadi:2005gi,:2008uu}.
This necessary homework might take even longer than the time we need to find new signals
(if we are lucky, of course!).
The Standard Model has some deficits, e.g. it does not contain a natural dark matter
candidate, it has no direct relation to gravity and it is sensitive to quadratic renormalisation effects.
Although the latter is in principle not a fundamental problem for a renormalisable 
theory like the Standard Model, a popular extension is supersymmetry (susy) which
would not only cure the problem above a sensible susy breaking scale 
of about 1 TeV, but would even allow to view the 
SM as a low energy limit of some string theory\footnote{If it is ever
possible to verify phenomenologically whether string theory is realised in nature
is of course a different question. It is allowed to be sceptical here.}.
Supersymmetric extensions of the Standard Model have been extensively studied
in the last years, for an overview see \cite{Martin:1997ns,Djouadi:2005gj,Allanach:2006fy}. 
Other extensions embed the SM in higher dimensions 
\cite{Sundrum:2005jf,Csaki:2005vy,Kribs:2006mq} 
or view the Higgs boson as a 
Goldstone boson to explain its relatively low mass, i.e. little Higgs models \cite{Schmaltz:2005ky}.
A further possibility is a strongly interacting theory which might
explain the mass of the W and Z-bosons even without a fundamental Higgs  boson.
There is of course another disturbing logical option: we see neither a hint
of the Higgs boson nor any sign of BSM physics. Although this would be a nightmare
for experimentalists it is actually an exciting option for theoreticians.
Such a scenario can only be realised if something "invisible" disguises
a low mass Higgs boson or if there are some strange interactions active in or beyond 
the  electroweak sector which fake a light Higgs boson in the precision measurements
by a yet unknown effect. 
Either we have to add an invisible sector to the Standard Model or we
need to question our quantum field theoretical description thereof, both would 
be truly exciting! 

\begin{figure}
\unitlength=1mm
\begin{picture}(120,65)
%\put(40,0){\includegraphics[width=7.cm]{ATLAS_signal_sig_30ifb.eps}}
\put(5,0){\includegraphics[width=6.5cm]{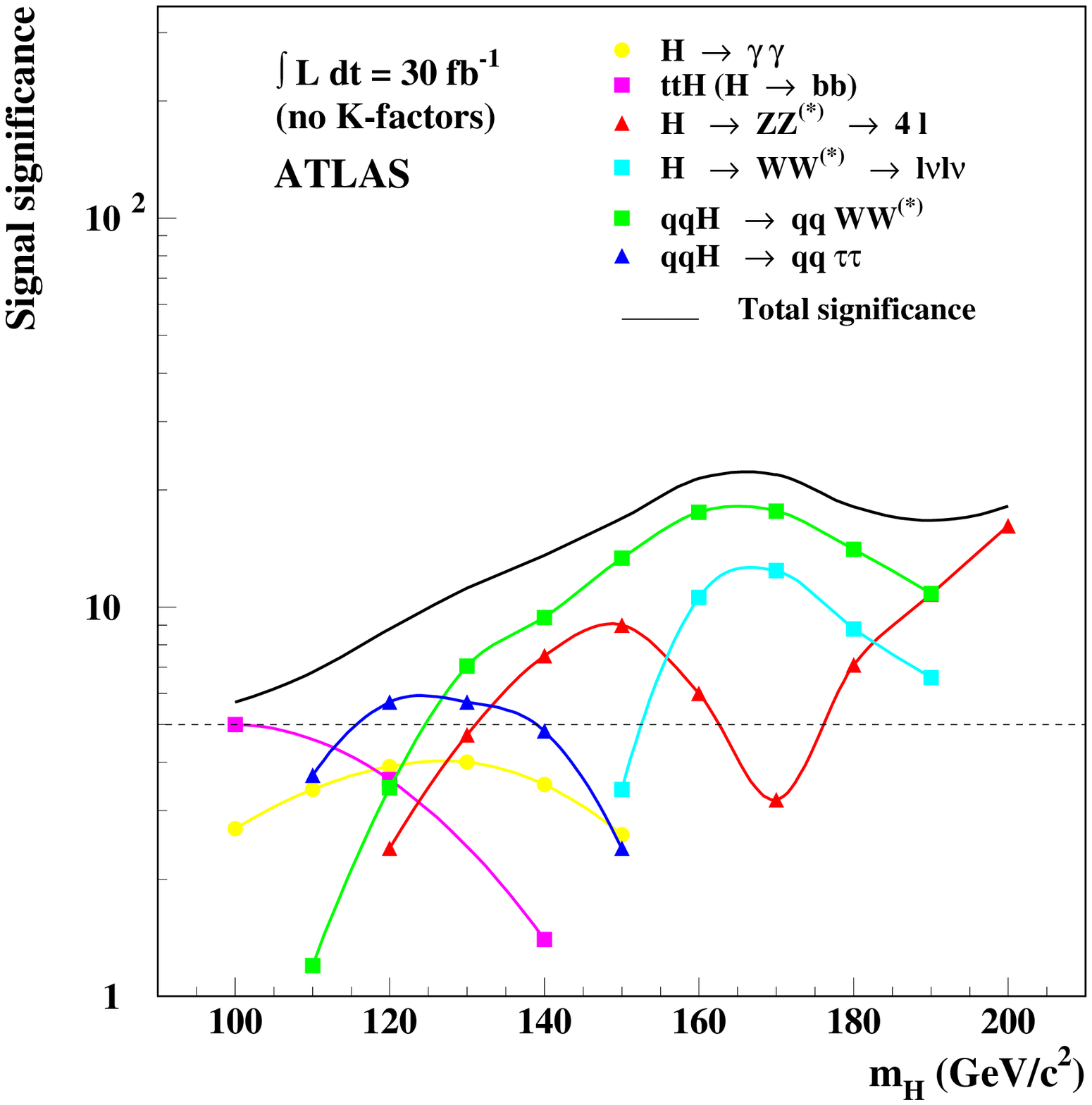}}
\put(75,0){\includegraphics[width=6.5cm]{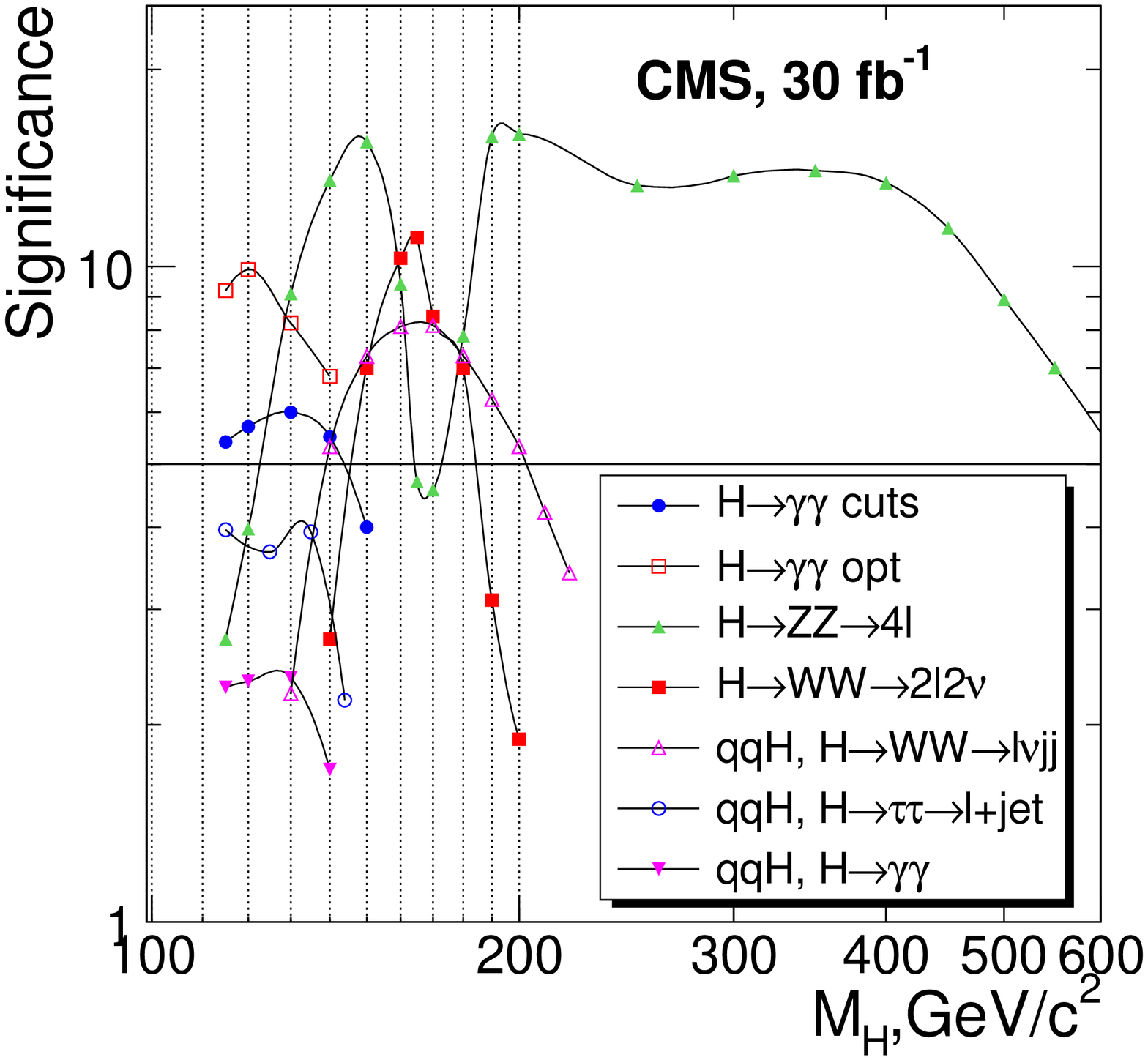}}
\end{picture}
\caption{The signal significance to find the Standard Model Higgs boson at the LHC
in various discovery channels for an integrated luminosity of 30 fb$^{-1}$ \cite{atlas:old,cms:old} 
(for reviews see \cite{Jakobs:2007zza,Buescher:2005re}).}\label{significance_plot}
\end{figure}

The discovery potential of the Higgs boson has been investigated thoroughly 
in the experimental studies of ATLAS and CMS  \cite{atlas:old,cms:old}
and signal significance plots like the one in 
Fig.~\ref{significance_plot} 
show the importance and relevance of different search channels depending on the Higgs mass. 
Note that in most experimental
studies K-factors, i.e. the effect of higher order corrections, have not been taken 
into account yet because higher order corrections to many background processes
were either not known or the computer codes of known calculations are not flexible 
enough. Higher order corrections
have been considered where available in recent experimental studies of the LHC 
physics potential \cite{Aad:2009wy}.

Preferably discoveries at the LHC should not depend on 
the theoretical status of theoretical simulations. For example the
diphoton decay of the Higgs boson should lead to a clear peak 
over the falling continuous background in the invariant mass distribution. 
This background should be measurable from the sidebands. 
Of course, a precise prediction is still needed for realistic studies
before measurement \cite{Binoth:1999qq,Bern:2002jx} and apart from that, who knows, 
whether the diphoton channel 
is not further contaminated by new physics?
Other discovery modes, e.g. the $H\to W^+W^-$ discovery mode \cite{Kauer:2000hi}, 
has no clear signal and background separation and background subtraction has to rely
on extrapolations of  background to signal regions which do depend on 
the precision of theoretical input. 

In what concerns the evaluation of leading order estimates for 
LHC cross sections based on tree-level matrix elements the situation is
quite satisfying as many tools 
have been developed in the pre-LHC era, e.g. {\tt Alpgen} \cite{Mangano:2002ea}, 
{\tt COMPHEP}  \cite{Pukhov:1999gg},  {\tt Amegic++}   
\cite{Krauss:2001iv}, {\tt COMIX}   \cite{Gleisberg:2008fv},  
{\tt FeynArts}/{\tt FormCalc}   \cite{Hahn:1998yk},  
{\tt GRACE}   \cite{Belanger:2003sd},
 {\tt HELAC} \cite{Kanaki:2000ey,Papadopoulos:2006mh,Cafarella:2007pc}, 
 {\tt Madgraph}/{\tt Madevent}   \cite{Maltoni:2002qb,Stelzer:1994ta}, 
 {\tt Whizard}   \cite{Kilian:2007gr},
and
most of them are publicly available. 
However, any partonic fixed order calculation is of limited use for experimental
studies which need realistic hadronic final states. This is the reason
why merging of partonic amplitudes with parton showers has been an active field recently. 
The standard Monte-Carlo tools which contain partonic showering and 
include a hadronisation model are currently {\tt Pythia} \cite{Sjostrand:2006za}, {\tt Herwig}
\cite{Bahr:2008pv} and {\tt Sherpa} \cite{Gleisberg:2008ta}. Evidently,
the more hard matrix elements are included (or better can be recursively evaluated) 
in these tools, the more accurate is the description of shapes of 
distributions and jet structure. In this respect  {\tt Sherpa}
is especially promising, as it is based on the 
matrix element generator {\tt Amegic++} \cite{Krauss:2001iv}.
On the other hand, the overall normalisation is predicted
only very unreliably as long as virtual higher order corrections are not taken into account.
This brings us to the main topic of the talk which is the computation
of these higher order corrections.  

\section{Framework for QCD computations}

The evaluation of production cross sections in hadronic collisions is based
on the factorisation property of QCD. Whenever the hard scale of the process is
significantly higher than the hadronisation scale of around one GeV
a differential cross section can be obtained by summing incoherently over the 
initial partons inside the hadron and  can be decomposed into long and short distant contributions.
\begin{eqnarray}
&& d\sigma(H_1H_2 \to \varphi_1 + \dots + \varphi_N + X) =
\sum\limits_{j,l} \int dx_1 dx_2 \;
f_{j/H_1}(x_1,\mu_F)\;  f_{l/H_2}(x_2,\mu_F) \nonumber\\ 
&& \qquad \times \, 
d\hat\sigma( \textrm{parton}_j(x_1P_1)+\textrm{parton}_l(x_2P_2)\to
\varphi_1 + \dots + \varphi_N, \alpha_s(\mu),\mu_F ) \nonumber
\end{eqnarray}
Here $\varphi_1 + \dots + \varphi_N$ stands for some partonic $N$-particle final state.
The parton distribution functions, $f_{j/H}(x,\mu_F^2)$, which stand for the probability to find a 
certain parton in an incoming hadron, have to be determined experimentally at a certain 
scale. Their scale dependence on the other hand is determined perturbatively 
through the DGLAP-evolution equations. 

The factorisation scale ($\mu_F$) dependence is stemming from the infrared (IR) 
structure of initial state
singularities. In contrast, the  renormalisation scale ($\mu_R$) dependence of 
the strong coupling constant is an ultra-violet (UV) effect. A hadronic cross section
in leading order (LO) perturbation theory is thus plagued by 
a logarithmic dependence on these scales. This dependence is 
aggravated when the number of coloured partons increases due to 
higher powers of $\alpha_s$. A LO computation in QCD has thus 
only limited predictive power. By including next-to-leading order (NLO)
corrections, the leading logarithmic dependence     
cancels and one obtains a much milder dependence on the unphysical scales.

The evaluation of NLO corrections contains a loop and 
real emission part. Schematically

   \unitlength=1mm
   \begin{picture}(100,20)
   \put(5,7){${\cal A}_{\textrm{LO}}$:}
   \put(17,4){\includegraphics[width=1.5cm]{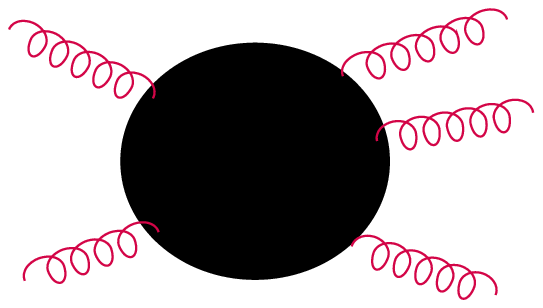}}
   \put(42,7){${\cal A}_{\textrm{NLO, virtual}}$:}
   \put(70,4){\includegraphics[width=1.5cm]{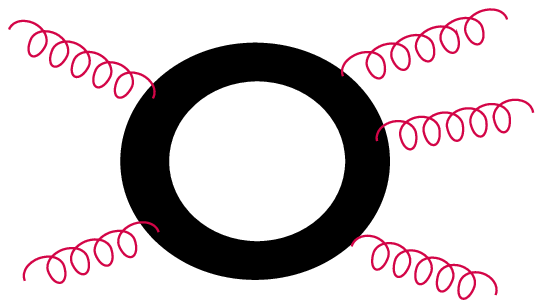}} 
   \put(95,7){${\cal A}_{\textrm{NLO, real}}$:}
   \put(120,4){\includegraphics[width=1.5cm]{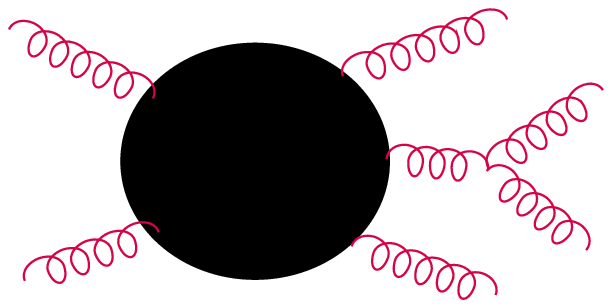}.}
   \end{picture}

The loop contribution has the same particle content and
kinematics as the LO contribution. The real-emission contribution
is tree-like but contains an additional parton. 
Apart from initial state collinear divergences, which
are absorbed by the parton distribution functions, 
the IR divergences cancel between the virtual corrections
and the integration of the real emission contributions over 
the soft/collinear phase space regions.  
In actual calculations the IR divergences are typically treated
by subtracting terms which have the same IR limit as the 
real emission corrections and do allow for the analytic
soft/collinear phase space
integration. Adding back the integrated terms
allows to represent  the $2\to N$ (NLO virtual) and $2\to N+1$ 
(NLO real) part as independently finite contributions, which is mandatory in 
numerical implementations. 

\begin{eqnarray}
\sigma &=& \sigma_{LO} + \sigma_{NLO} \nonumber \\
\sigma_{LO} &=& \int dPS_N \;\frac{1}{2s}\; %\mathcal{O}_N(\{p_j\}) 
\;\; |{\cal A}_{\textrm{LO}}|^2 \nonumber \\
\sigma_{NLO} &=&  \int dPS_N \;\frac{1}{2s}\;\alpha_s\;
  %\,\mathcal{O}_N(\{p_j\})
  \;\Bigl[ {\cal A}_{\textrm{LO}}{\cal A}_{\textrm{NLO, virt.}}^\dagger 
+ {\cal A}_{\textrm{LO}}^\dagger{\cal A}_{\textrm{NLO, virt.}} 
+ \sum_j \int dPS_{1,j} {\cal D}_j \Bigr]  \nonumber\\
&& \quad+\int dPS_{N+1} \;\frac{1}{2s}\;\alpha_s\;
% \,\mathcal{O}_{N+1}(\{p_j\})
\;\, \Bigl[|{\cal A}_{\textrm{NLO, real}}|^2 - \sum_j {\cal D}_j \Bigr] 
\end{eqnarray}
The most wildly used variant of these subtraction methods is the so-called
dipole subtraction formalism developed by Catani and Seymour \cite{Catani:1996vz}
which has been generalised to the case of massive partons in \cite{Catani:2002hc}.
We see that the ingredients of a NLO QCD calculation consist  of
(i) tree level amplitudes, (ii) one-loop amplitudes and (iii)
dipole subtractions terms, ${\cal D}_j$. As was pointed out above, 
many automated tools are available for the evaluation of the tree-level contribution. 
Moreover, different groups have recently implemented
the dipole formalism in a process independent way 
\cite{Seymour:2008mu,Gleisberg:2007md,Hasegawa:2008ae,Frederix:2008hu}, and
public codes exist which help to perform this labour intensive part
of an NLO computation. The real bottleneck of an automated approach
for NLO QCD computations is thus the evaluation of the one-loop
matrix elements. This will be the main subject of the remainder
of this talk. 

\section{New methods to compute one-loop amplitudes}

\subsection{Overview of achieved results}
The complexity of loop calculations grows rapidly 
with the number of external legs.
What concerns computations with two initial and two final
particles, basically all relevant LHC processes are evaluated
meanwhile at NLO in QCD and public codes, like for example 
{\tt MCFM} \cite{Campbell:2000bg,mcfm_page}, the {\tt PHOX} 
programs for observables including photons\cite{phox_page}, 
and {\tt VBF@NLO} for weak boson fusion processes \cite{Arnold:2008rz}, 
are available for experimental studies on the partonic level. 
Methods to combine fixed order calculations  with parton
showers have been worked out and implemented for many important
processes, see for example {\tt MC@NLO} \cite{Frixione:2008ym}, 
{\tt POWHEG} \cite{Frixione:2007vw}, 
{\tt GRACE} \cite{Fujimoto:2008zz} and \cite{Nagy:2005aa,Lavesson:2008ah}. 
Also the {\tt SANC} collaboration 
has promoted their framework to include QCD processes
and progress has also been reported during this 
conference \cite{Andonov:2003xe,Andonov:2007zz,Bardin:2007zz}.
A relevant gap of $2 \to 2 $ processes has been closed recently.
The off-shell vector boson pair production cross sections
for $gg\to W^*W^*,\;\gamma^*\gamma^*,\;\gamma^*Z^*,\;Z^*Z^*$
are available now as public codes {\tt GG2WW} \cite{Binoth:2005ua,Binoth:2006mf}
and {\tt GG2ZZ} \cite{Binoth:2008pr}. 
These processes describe a relevant background component 
to $H\to WW^*/ZZ^*$ also below threshold. 
These processes are induced by loop diagrams but are 
conceptually of LO type. Technically one has 
to square one-loop amplitudes which is numerically more delicate
than computing interference terms between tree and loop
contributions. 

Adding an additional final state particle increases the
complexity of one-loop computations already considerably, and
before 2005 only a limited number of $2\to 3$ processes had been
evaluated and implemented in computer programs.
NLO predictions have been existing for 
$pp\to jjj$ \cite{Nagy:2003tz},
$pp \to \gamma\gamma j$ \cite{deFlorian:1999tp,DelDuca:2003uz},
$pp\to Hjj$ in gluon fusion \cite{Campbell:2006xx} and 
weak boson fusion (WBF) \cite{Han:1991ia,Figy:2003nv}, 
and $pp\to Ht\bar{t}$ \cite{Beenakker:2002nc,Dawson:2003hv}.
Between 2005 and this date most of the relevant 
$2\to 3$ processes for LHC have been completed
and the list of accomplished tasks contains 
also the processes
$pp \to HHH$ \cite{Plehn:2005nk,Binoth:2006ym},
$pp \to VVjj$ in WBF\footnote{The six external particles
are connected via at most 5-point one-loop topologies.} \cite{Jager:2006zc,Jager:2006cp,Bozzi:2007ur},
$pp \to ZZZ$ \cite{Lazopoulos:2007ix}, 
$pp \to t\bar{t} j$\cite{Dittmaier:2007wz,Dittmaier:2008uj},
$pp \to WWj$ \cite{Dittmaier:2007th,Campbell:2007ev}, $pp \to VVV$ \cite{Binoth:2008kt,Hankele:2007sb}
and $pp \to t\bar{t}Z, b\bar{b}Z$ \cite{Lazopoulos:2007bv,Lazopoulos:2008de,FebresCordero:2008ci}.
For the important $pp\to Hjj$ process also  electroweak \cite{Ciccolini:2007ec,Ciccolini:2007jr}
and sub-leading interference terms have been evaluated
\cite{Weber:2006au,Andersen:2006ag,Andersen:2007mp,Bredenstein:2008tm}.  
A discussion of the phenomenological importance of the different
processes can be found in \cite{Bern:2008ef}.
Note that the evaluation of the relevant amplitudes is
only about half of the cake. The implementation of 
a full NLO cross section into a reliable computer code 
is also a considerable effort due to numerical issues which will 
be discussed below. Further note that most of the corresponding codes are
not public. 

The time-scale of all these calculations were of the
order of person-years and it is no wonder
that only a very limited number of results which deal with NLO $2 \to 4$ LHC 
processes can be found in the 
literature\cite{Bredenstein:2008zb,Binoth:2008gx,Ellis:2008qc,Berger:2009zg,Reiter:2009kb}. 
In the remainder of this section I will focus 
on the recent developments in this direction
and report on progress which has been achieved by
different groups and different methods recently.
The two main approaches are based on either Feynman 
diagrams or on unitarity cuts. Before discussing these two
approaches separately let us shortly comment on two 
important commonly used techniques.

\subsection{Colour and helicity management}

When evaluating gauge boson amplitudes or dealing with massless
fermions it is very useful to work with spinor helicity 
methods, for an introduction see \cite{Dixon:1996wi}.
It is well known that tree-level amplitudes are most efficiently
represented in this way. The same is true for loop amplitudes.
A massless Dirac spinor is already defined by two helicity states $| k^\pm \rangle$ defined by
\begin{eqnarray}
k \!\!\!/| k^\pm \rangle = 0 \quad , \quad | k^\pm \rangle = \Pi^\pm u(k) \quad , 
\quad  \langle k^\pm | = \bar{v}(k) \Pi^\mp  \quad , \quad 
\Pi^\pm = \frac{1}{2}(1\pm \gamma_5) \; .\nonumber
\end{eqnarray}
Helicity amplitudes can be written  in terms of spinor products which are
complex numbers:
\begin{eqnarray} 
\langle k q \rangle = \langle k^- | q^+ \rangle &,& [ k q ] = \langle k^+ | q^- \rangle   \; .\nonumber
\end{eqnarray}
Massless gauge bosons like gluons and photons can be expressed using
the same building blocks 
\begin{eqnarray}\label{eq.sp}
\varepsilon_{\mu}^+(k,q) = 
\frac{1}{\sqrt{2}}\frac{\langle q^- |\gamma_\mu  | k^-\rangle}{ \langle q k\rangle } &,&
\varepsilon_{\mu}^-(k,q) = \frac{1}{\sqrt{2}}\frac{\langle q^+ | \gamma_\mu  | k^+\rangle}{ [kq] }\; .\nonumber
\end{eqnarray}
By construction one works with the two physical degrees of freedom.
The auxiliary vector $q$ defines an axial gauge. In full amplitudes
its dependence drops out due to gauge invariance.
The generalisation to massive particles is also well-understood.
A public implementation of spinor helicity methods can be found in \cite{Maitre:2007jq}.

Once the helicity amplitudes are evaluated, the squared amplitude
can be obtained numerically. If there are $K$ helicity amplitudes
labelled by $\{\lambda_j\}$ one only has to evaluate $K$ products
\begin{equation}
{\cal A} = \sum\limits_{\lambda_j} {\cal A}^{\{\lambda_j\}}
\Rightarrow |{\cal A}|^2 
= \sum\limits_{\lambda_j}  {\cal A}^{\{\lambda_j\}\,*} {\cal A}^{\{\lambda_j\}} \quad ,
\end{equation}
otherwise $K^2$ terms have to be evaluated.
%Not using helicity amplitudes needs the evaluation of $K^2$ terms.
To give an example, the $N$-gluon amplitude can be decomposed into 
$2^N$ helicity amplitudes. Due to parity invariance only  $2^{N-1}$ have 
to be evaluated. Nonetheless we see an exponential growth of terms in N.
Another source for exponential growth in N-point amplitudes
is the colour structure.
A widely used colour decomposition is based on the following
two elementary rules valid for an $SU(N_C)$ algebra with
generators $T^a$ in the fundamental representation:
\begin{eqnarray}
i f^{abc} T^c_{ik} &=& T^a_{ij} T^b_{jk} - T^b_{ij} T^a_{jk} \nonumber \\
T^a_{ik} T^a_{jl}  &=& \frac{1}{2} \Bigl( \delta^i_l \delta^j_k - \frac{1}{N_C}  \delta^i_k \delta^j_l  \Bigr)  \nonumber
\end{eqnarray}
Any colour structure can be  mapped to simple Kronecker deltas
which indicate the colour flow. This colour flow decomposition, described
for example in
\cite{Kanaki:2000ms,Maltoni:2002mq},  is going back to $\mbox{}'$tHooft 
who applied a double line notation for gluons in the context of the $1/N_C$ expansion.
%maps to colour basis (\textcolor{blue}{N = \# gluons $\leftrightarrow$ \#quark lines})
Any amplitude with $N_F$ fermion lines and $N_G$ gluons can be decomposed 
in terms of $N!=(N_F+N_G)!$ products of colour flow lines. The different
elements are labelled by the permutation group  $S_N$ acting on 
colour line indices. 
\begin{eqnarray}
{\cal A} = \sum\limits_{\sigma \in S_N}  {\cal A}_\sigma |c_\sigma \rangle &,&
| c_\sigma \rangle = \delta_{i_1}^{ j_{\sigma(1)} }  \delta_{i_2}^{ j_{\sigma(2)} } \dots \delta_{i_N}^{ j_{\sigma(N)} }
\end{eqnarray}
For gluon amplitudes this leads to an over-counting of independent colour states,
as some singlet contributions are actually forbidden. The 
N-gluon amplitude has $(N-2)!$ independent colour states.
Still, asymptotically we notice a factorial growth of terms
for large $N$.
A way to fight the exponential growth both in helicity and colour
components may be to resort to Monte Carlo sampling over
these degrees of freedom, see for example \cite{Papadopoulos:2005ky}.

\subsection{The unitarity based approach}

The idea to use the analytic structure of scattering amplitudes
to determine their explicit form is very old \cite{Cutkosky:1960sp,analytic_s_mat}.
The well known non-linear relation between the transfer matrix, $T$, implied by 
the unitarity of the S-matrix, $S=1+iT$,
\begin{equation}
S^\dagger S = 1 \Rightarrow 2 \;\textrm{Im}(T) = T^\dagger T \; ,
\end{equation}
leads to a relation between the imaginary part of
one-loop amplitudes and sewed tree-level diagrams,
schematically:

\unitlength=1mm 
\begin{picture}(100,15)
\put(33,6){\textrm{Im} ${\cal A}_{1-\textrm{loop}} \sim \sum\limits_C\int d\textrm{PS}_C$}
\put(72,-3){\includegraphics[height=20mm]{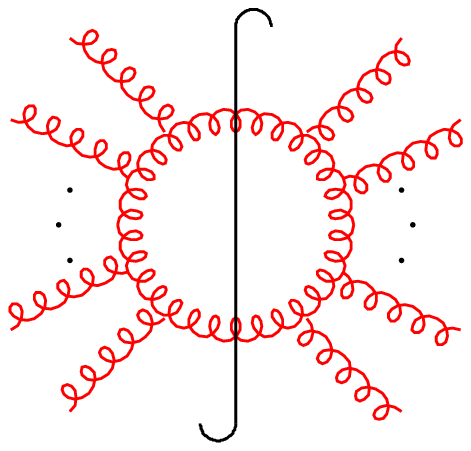}}
%\put(70,0){$\vert_$}
\end{picture}

Here the right hand side stands for the sum over all two-particle cuts.
The latter correspond to squared tree-level amplitudes integrated
over the respective two-particle phase space.
Using the standard Feynman diagrammatic approach it can be shown 
that any one-loop amplitude with massless internal particles
can be decomposed in terms of known scalar integrals with two to four
external legs\footnote{If masses are present one also has to consider
one-point integrals.} $I_{k=2,3,4}$,  and a rational
part, $\mathcal{R}$, which does not contribute to the imaginary part of the amplitude.
The coefficients of the integrals, $C_k$, and the rational term
are rational polynomials in terms of spinor products, see Eq.~(\ref{eq.sp}),
and/or Mandelstam variables and masses.
The imaginary parts of the different scalar integrals
can be uniquely attributed to a given integral. Subsequently
unitarity implies that the knowledge of the imaginary 
part of the amplitude defines the coefficients
of the scalar integrals.   
\begin{eqnarray}\label{avsints}
{\cal A}_{1-\textrm{loop}} = \sum_{k=2,3,4} C_k I_k + \mathcal{R} & \Rightarrow &
\textrm{Im}\, {\cal A}_{1-\textrm{loop}} = \sum_{k=2,3,4}  C_k\, \textrm{Im}( I_k ) 
\end{eqnarray}
Note that there are many different scalar two-, three- and four-point functions
present in a given process. This is not indicated in this schematic equation.
We see that tree amplitudes and phase space integrals in D=4 dimensions  
already fix a large part of the amplitude without ever being forced
to evaluate any one-loop diagram. By turning the argument around we learn
that one-loop diagrams are to a great extent determined by the pole part
of the internal propagators which are nothing but cut propagators. 
The rational part is 
actually stemming from the UV behaviour of the one-loop amplitude,
it comes from terms $(D-4)\, I_2^D \sim 1$. In amplitudes with an improved
UV behaviour, such as susy amplitudes, these terms are absent,
and this explains why these amplitudes are fully determined by
the D=4 cut structure. This fact has seen many applications in the 
last years, see e.g. \cite{Bern:2007dw}.
As the rational part is not
determined by the D=4 cut structure, one has to resort to more
sophisticated methods, like for example on-shell recursion relations \cite{Berger:2006ci}
or D-dimensional unitarity \cite{Anastasiou:2006jv}. 

This method has been developed and applied to a great
extent by Bern, Dixon, Dunbar and Kosower in the 90s  
\cite{Bern:1994cg,Bern:1994zx}, for a recent review see \cite{Bern:2007dw}.
It has seen a renaissance after the Santa Barbara workshop in 2003
where Witten pointed out a relation between certain string theories
and QCD amplitudes using
 twistor methods \cite{Witten:2003nn}. The revived interest lead to new efficient
recursive evaluation procedures for multi-parton amplitudes 
\cite{Britto:2004nc,Brandhuber:2005jw,Britto:2006sj}. 
A main feature of recent variants of the method is that multiple cuts are used 
to determine integral coefficients, which goes under the name of {\it generalised unitarity}
\cite{Britto:2004nc,Brandhuber:2005jw}.
\begin{figure}
\unitlength=1mm 
\begin{picture}(100,35)
\put(35,0){\includegraphics[height=3.5cm]{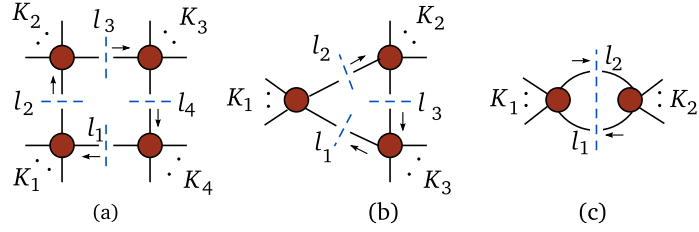}}
\end{picture}
\caption{Multiple cuts can be used to fix integral coefficients
of amplitudes.}\label{cut_fig}
\end{figure}
Cutting four lines in an N-point topology amounts to putting the
corresponding four propagators on-shell. This procedure fixes
the associated loop momentum completely  and the 
coefficient of the related box diagram, see Fig.~\ref{cut_fig}(a), is given as
a product of tree diagrams.
\begin{eqnarray}
\textrm{Fig.~\ref{cut_fig}(a):} \quad \Rightarrow \quad C_4 = \frac{1}{2} \sum\limits_{\sigma=\pm}
  A^{(1)}_\textrm{tree} \, A^{(2)}_\textrm{tree} \,A^{(3)}_\textrm{tree} \,A^{(4)}_\textrm{tree} \nonumber
 \end{eqnarray}
One has to sum over the two complex solutions of  
quadratic on-shell conditions. This fixes the loop momentum.
An important point is that the tree level amplitudes already incorporate 
gauge invariance manifestly. In Feynman diagram computations
many graphs have to be combined to
result in a gauge invariant expression. 
The power of this method has been demonstrated by the analytic evaluation
of the cut-con\-struct\-able part of the six-gluon amplitude \cite{Britto:2006sj}. 
Combined efforts of many groups were needed to compute the different
pieces of this amplitude, see \cite{Dunbar:2008zz,Dunbar:2009uk}
for a collection of all formulae. The rational 
part has been first provided by a Feynman diagrammatic computation 
\cite{Xiao:2006vt,Su:2006vs,Xiao:2006vr}.
Note that the six-gluon one-loop amplitude is part of a full NLO computation
of four jet production at the LHC and thus of phenomenological interest.
There exists also a very compact result of the six-photon amplitude
which was obtained using cutting methods \cite{Binoth:2007ca,Ossola:2007bb}. As has been
shown in \cite{Binoth:2006hk}, the rational part is zero in this case.

The unitarity method, using multiple cuts, has recently been implemented 
in a numerical code by the  {\tt Blackhat} collaboration \cite{Berger:2008sj,Berger:2008ag}
and by Ellis et. al. \cite{Ellis:2007br} (discussed below). 
To this moment the {\tt Blackhat} collaboration 
has provided two remarkable applications, the colour ordered part of the 
(N=6,7,8)-gluon amplitudes in a special helicity configuration, ${\cal A}^{--++\dots+}$
and the leading colour contribution of $q\bar{q} \to Vggg$ 
(this includes  crossing related amplitudes) \cite{Berger:2008sz,Berger:2009zg}.
 
\begin{figure}
\begin{picture}(100,50)
\put(10,0){\includegraphics[height=5.0cm]{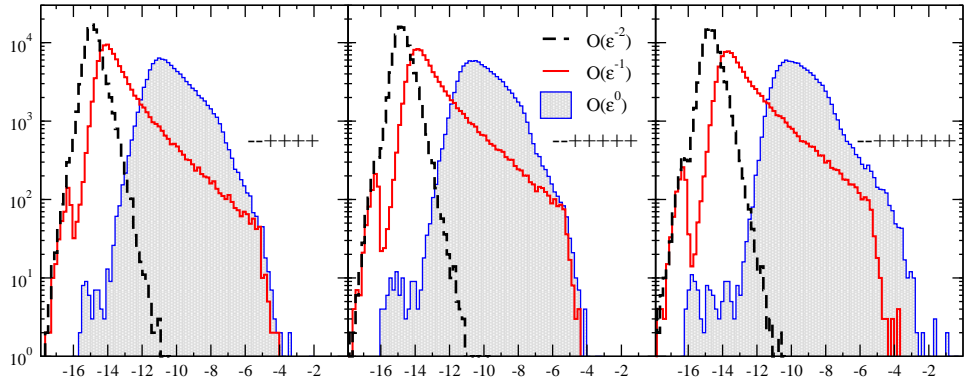}}
\end{picture}
\caption{The histgrams of the evaluated events are plotted over the logarithmic 
relative error $\log(|{\cal A}_{\textrm{numeric}} 
- {\cal A}_{\textrm{target}})|/|{\cal A}_{\textrm{target}}|)$
of the pole and finite parts of the 
MHV N=6,7,8-amplitudes. The shaded region (blue) 
relates to the finite contribution,
The unshaded full line region (red) to the $1/\epsilon$ pole part and the dashed
to the double-pole part \cite{Berger:2008sj}.}\label{blackhat_plot}
\end{figure}

Important information
on the numerical behaviour of the method could be gained in \cite{Berger:2008sz}.
The large number of algebraic operations in these complex amplitudes
typically lead to round up errors in certain exceptional 
phase space regions. Any automated method has to come up with
a reliable fail-safe procedure if the numerical cancellations
are spoilt by the finite number of computed digits. In Fig.~\ref{blackhat_plot}
a comparison of the MHV 6,7,8-gluon amplitudes between the numerical implementation and 
the known analytical result is shown.       
The logarithmic relative error, 
$\log(|{\cal A}_{\textrm{numeric}} 
- {\cal A}_{\textrm{target}})|/|{\cal A}_{\textrm{target}}|)$, is plotted for the double, 
single-pole and the finite part of these amplitudes for 100.000 phase space points
in Fig.~\ref{blackhat_plot}.
While the bulk of all points is evaluated reliably
using standard double precision, it can be observed 
that the distribution has a tail 
which for a few percent of phase space points indicates 
a precision loss of almost all digits. Such a situation typically
leads to an unstable behaviour of adaptive Monte Carlo event generators,
which tend to sample points exactly in regions which induce a large variance
of the result. In the given plot one observes that the tail is cut at a precision
of around $10^{-3}$ to $10^{-4}$. Beyond that point multi-precision libraries  \cite{bailey_talk}
are used to avoid a dangerous loss of precision.

\begin{figure}
\unitlength=1mm 
\begin{picture}(100,55)
\put(35,0){\includegraphics[height=5.5cm]{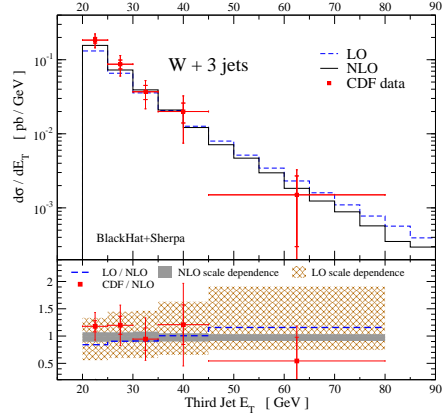}}
\end{picture}
\caption{The leading colour NLO prediction for $p\bar{p} \to Vjjj$ 
compared to Tevatron data. The leading colour prediction is expected 
to match the full result within a few percent. As expected the error 
bands due to scale variations reduce drastically when moving from LO (brown)
to NLO (grey) \cite{Berger:2009zg}.}\label{wjjj}
\end{figure}

The evaluation of the loop amplitude to 
$q\bar{q}'\to Wggg$ is the necessary ingredient for a  prediction 
of the important Standard Model background $pp\to Vjjj$.
Such multi-jet plus lepton plus missing energy signals 
occur for example in supersymmetric extensions of the 
Standard Model. 
The {\tt Blackhat} collaboration has confronted their  
leading colour prediction for $p\bar{p} \to Vjjj$ \cite{Berger:2009zg} with Tevatron data, see
Fig.~\ref{wjjj}.
As expected the stability of the prediction under scale variations 
is largely improved.
The leading colour approximation is expected to be phenomenologically justified
within a few percent for Tevatron kinematics.
Technically this approximation simplifies the calculation considerably, 
as many topologies are suppressed and can be neglected in this way.

\bigskip

In 2006 Ossola, Papadopoulos and Pittau (OPP) have proposed another unitarity approach
which was based on the question how to reconstruct one-loop amplitudes
from the integrand of Feynman integrals \cite{Ossola:2006us}. Let us consider a sub-amplitude
where all external particles are ordered. The full amplitude can be obtained
by adding up different permutations.
The leg-ordered amplitude can be obtained after summing the corresponding
Feynman diagrams written such that the sum is under the
integral and has a common denominator of propagators $D_1\dots D_N$
with $D_j=(k-r_j)^2 - m_j^2$.
\begin{eqnarray} 
{\cal A}_{1,\dots, N} = \sum {\cal G}_{1,\dots, N} \sim \int d^Dk \frac{{\cal N}(k)}{D_1\dots D_N} \nonumber
\end{eqnarray} 

The fact that general one-loop amplitudes can be written in terms
of at most 4-point scalar integrals implies that the numerator 
can be written schematically in the following way 
\begin{eqnarray}\label{opp_num}
{\cal N}(k) &\sim& 
  \sum_{boxes}     [ C_4 + \tilde{C}_4(k) ] \prod\limits_{j \in \!\!\!\!/ box}^{N} D_j 
+ \sum_{triangles} [ C_3 + \tilde{C}_3(k) ] \prod\limits_{j \in \!\!\!\!/ triangles}^{N} D_j \nonumber \\
&& + \sum_{bubbles}   [ C_2 + \tilde{C}_2(k) ] \prod\limits_{j \in \!\!\!\!/ bubbles}^{N} D_j + \dots 
\end{eqnarray}
In this expression the coefficients $C_4$, $C_3$, $C_2,\dots$, correspond to the
actual amplitude coefficients of the various box, triangle and bubble integrals 
defined above, see Eq.~(\ref{avsints}). While these coefficients
are independent  of the loop momentum, the numerator function also contains
spurious k-dependent pieces,  $\tilde{C}_4$, $\tilde{C}_3$, $\tilde{C}_2,\dots$,
which have to vanish upon integration over the loop momentum $k$.
OPP asked the question how these spurious coefficients can be 
defined {\em without} doing the actual integration.
They concluded that one simply has to evaluate the on-shell conditions 
of the various propagators which is in one-to-one relation to the unitarity
cut methods using multiple cuts. 
This approach defines the coefficients
of the terms $\tilde{C}_4$ ,$\tilde{C}_3$, $\tilde{C}_2,\dots$,
as polynomials in $k$. The coefficients of the box-, triangle-, bubble terms 
are defined by respectively 2,7,9 $k$-polynomial coefficients. OPP proposed to extract
these coefficients by numerical interpolation of the various
polynomials in $k$ \cite{Ossola:2006us}. They also define a procedure to
access the rational part of the amplitudes \cite{Ossola:2008xq,Draggiotis:2009yb}.
This method works also for individual one-loop Feynman diagrams and 
it has been implemented in a public computer code, {\tt CutTools} \cite{Ossola:2007ax}.
%As it is a numerical method one is again plagued by 
The method has been applied successfully to the evaluation of
six-photon amplitudes with massless and massive internal fermions \cite{Ossola:2007bb}. 
%\cite{Bernicot:2007hs}.
%The massless case shows perfect agreement with known results, the
%massive case has not yet been evaluated by other methods.
The method has  also been applied  to the QCD corrections of 
triple vector boson production at the LHC. Here the amplitude has been 
integrated to obtain the total and
differential cross sections for relevant observables \cite{Binoth:2008kt}.   

\bigskip

A variant of these numerical unitarity based methods has also been 
investigated in \cite{Ellis:2007br}
and has been extended to a D-dimensional approach \cite{Ellis:2008kd,Ellis:2008ir}. 
In the latter case also 5-point functions have to be included in
the function basis. The key observation is that 
by adding higher dimensional information 
the rational part of the amplitude is also determined.
A D-dimensional numerator function of a leg ordered
amplitude is of the form
\begin{eqnarray}
{\cal N}_D &=& {\cal N}_4 + (D-4) {\cal N}_{D-4} 
           = \sum_{pentagons}  [ C_5 + \tilde{C}_5(k) ] 
\prod\limits_{j \in \!\!\!/ pentagon}^{N} D_j 
  + \textrm{[r.h.s. of Eq.~(\ref{opp_num})]} \nonumber
\end{eqnarray}
This approach guarantees that the evaluation of the rational part 
is now on the same footing as the  cut constructable part.
The method relies on the evaluation of amplitudes in
dimensions different from four. It can be viewed as a D-dimensional generalisation
of the OPP approach. The authors observed that the knowledge of
tree amplitudes evaluated for complex momenta which are
defined through the on-shell conditions of propagators,
or equivalently through multiple cuts, defines the complexity
of the algorithm modulo the number of different possibilities to 
cut a given N-point topology. The latter scales like $\sim N^5$, as
any set of five propagators have to be put on-shell
in a given diagram. The recursive evaluation of tree-amplitudes scales
approximately like $\sim N^4$. Thus it can be expected that 
a colour ordered multi-parton amplitude can be evaluated by such a
polynomial complexity algorithm, or short ${\cal P}$-algorithm,  \cite{Ellis:2008kd,Ellis:2008ir}.
As was pointed out above, for full amplitudes one needs to sum over
different helicities and colour structures which unavoidably 
turns on an exponential growth and thus one ends up with an exponential or
${\cal E}$-algorithm for full amplitudes.
Again Monte-Carlo sampling over colour and helicities might 
help to single out the relevant contributions
numerically in a more efficient way. If a polynomial algorithm
can be realised in this way remains to be shown.

The potential of this variant of the unitarity method has been 
demonstrated by evaluating single phase space points of different helicity amplitudes of
the colour ordered $N$-gluon amplitude for up to $N=20$ legs \cite{Giele:2008bc}.
Sampling over a large number of phase space points shows a 
good numerical performance concerning speed.  

\begin{figure}
\begin{picture}(100,45)
\put(35,45){\includegraphics[height=70mm,angle=270]{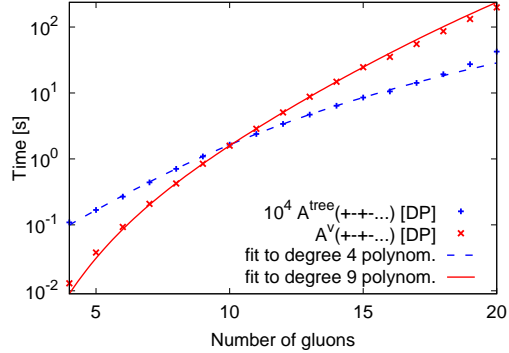}}
\end{picture}
\caption{Evaluation time of colour ordered $N$-gluon helicity amplitudes. 
The dependence of tree- and one-loop amplitudes on $N$ confirms
the expectation of a polynomial growth proportional to
$\sim N^4$ and $N^9$ respectively \cite{Giele:2008bc}.}\label{giele_plot}
\end{figure}

In Fig.~\ref{giele_plot} the polynomial behaviour
of tree and one-loop amplitudes is shown as a function of $N$.
The polynomial behaviour of the tree, $\sim N^4$, and the one-loop
amplitudes, $\sim N^9$, is confirmed for the multi-gluon amplitudes.
At the moment it seems not feasible to produce  such a plot based on
Feynman diagrams ever, as gauge cancellations and colour decomposition
are not manifest and one needs the evaluation of the order
of $N!$ form factors.
An independent implementation of the same method and amplitude,
confirming the result,
has been presented recently \cite{Lazopoulos:2008ex,Winter:2009kd}.
Other applications of the method are the evaluation of the
amplitudes $gggt\bar{t} \to 0$ \cite{Ellis:2008ir} and the leading
colour contributions to $q\bar{q}Wggg$,   $q\bar{q}q'\bar{q}'Wg$ 
\cite{Ellis:2008qc,Ellis:2009zw}.  The latter are also implemented in the {\tt Rocket} code
\cite{Giele:2008bc} which handles numerically problematic phase space points by 
using higher precision libraries. Again, ultimately multi-precision libraries are in use 
to guarantee numerical reliability \cite{bailey_talk}.

From the fast progress made in the last two years it can be expected 
that further cross section calculations for LHC processes 
with four or more partons/particles in the final state will
become available using the discussed unitarity based methods.   
Note that apart from the multiple cut methods described so far,
also single cut techniques are under development \cite{Catani:2008xa}.

\subsection{The Feynman diagrammatic approach}

The traditional method of performing amplitude calculations
is by starting from Feynman diagrams.
It is based on representing
a one-loop amplitude as a sum of  diagrams sorted by colour structures
\begin{eqnarray}
{\cal A}^{{c},{\lambda}}({p_j,m_j}) &=& \sum\limits_{\{ c_i\},\alpha}  f^{\{c_i\}}{\cal G}_\alpha^{\{\lambda\}}\nonumber \\
{\cal G}_\alpha^{\{\lambda\}} &=& \int\frac{d^Dk}{i\pi^{D/2}}
\frac{ {\cal N}^{\{\lambda\}} }{ D_1\dots D_N }
= \sum\limits_{R}  {\cal N}^{\{\lambda\}}_{\mu_1,\dots,\mu_R}\; I_N^{\mu_1\dots\mu_R}({p_j,m_j}) \nonumber\\
I_N^{\mu_1\dots\mu_R}({p_j,m_j}) &=& \int\frac{d^Dk}{i\pi^{D/2}}
\frac{ k^{\mu_1} \dots k^{\mu_R} }{ D_1 \dots D_N }
\;, \;  D_j = (k-r_j)^2-m_j^2 ,\; r_j=p_1\dots +p_j\; .
\end{eqnarray}
The kinematical functions can be represented by tensor one-loop integrals. 
The latter can be evaluated recursively in momentum space using
the Passarino-Veltman method \cite{Passarino:1978jh,Denner:1991kt}.
Alternatively one can map the momentum integrals to Feynman parameter
integrals which also can be reduced recursively 
\cite{Davydychev:1991va,Bern:1993kr,Tarasov:1996br,Duplancic:2003tv,Giele:2004iy}. 
\begin{eqnarray} 
I_{N}^{\mu_1\dots \mu_R} &=& \sum \tau^{\mu_1\dots \mu_R}(r_{j_1},\dots,r_{j_r},g^m) \:
                                  I_{N}^{D+2m}(j_1,\dots,j_r) \nonumber \\
I_N^{D}(j_1,\dots ,j_r) &=&  (-1)^N\Gamma(N-\frac{D}{2}) 
\int_{0}^{\infty} d^Nz\,\delta(1-\sum\limits_{l=1}^N z_{l})\,
\frac{z_{j_1}\dots z_{j_r}}{(-\frac{1}{2}z\cdot {\cal S}\cdot z )^{N-D/2}} \nonumber \\
{\cal S}_{ij} &=&  (r_i-r_j)^2-m_i^2-m_j^2 \; \nonumber .
\end{eqnarray}
The end-point of the different recursion algorithms
are scalar integrals with no numerator structure. Public program libraries
exist to evaluate the 
latter \cite{Ellis:2007qk,vanOldenborgh:1990yc,vanHameren:2005ed,Binoth:2008uq,Hahn:1999mt}.

Over the years quite a few groups have gathered a lot of experience
in efficiently implementing Feynman diagram calculations. 
Cross section calculations, with up to five external particles,
can generally be mastered now, as is demonstrated by the long list
of accomplished tasks discussed above.
For $2\to 4$ processes only a view NLO computations have been accomplished
using Feynman diagrams,
$e^+e^- \to f_1 \bar{f}_1 f_2 \bar{f}_2$ \cite{Denner:2005fg},
$e^+e^- \to HH\nu\nu$ \cite{Boudjema:2005rk}, and 
$\gamma \gamma \to b\bar{b} t\bar{t}$ \cite{Lei:2007rv}.
Processes of this complexity relevant for LHC phenomenology are 
under construction right now. Progress on the computation of the important LHC process
$pp \to b\bar{b} t\bar{t}$ was reported in 2008 \cite{Bredenstein:2008zb},
this group presented the full cross section computation of the
quark induced subprocess $q\bar{q} \to b\bar{b} t\bar{t}$ at NLO
in $\alpha_s$. The amplitude can be written in terms of rank three
6-point tensor integrals
\begin{eqnarray}
{\cal A}(q\bar{q} \to b\bar{b}t\bar{t}) \sim \sum C_{j_1j_2j_3} I_{N\leq 6}^{j_1j_2j_3} \; .
\end{eqnarray}
In this  calculation the algorithm described 
in \cite{Denner:2005nn} was used for the tensor reduction of 
5- and 6-point integrals, otherwise Passarino-Veltman reduction was used. 
\begin{figure}
\begin{picture}(120,50)
\put(20,0){\includegraphics[height=50mm,angle=0]{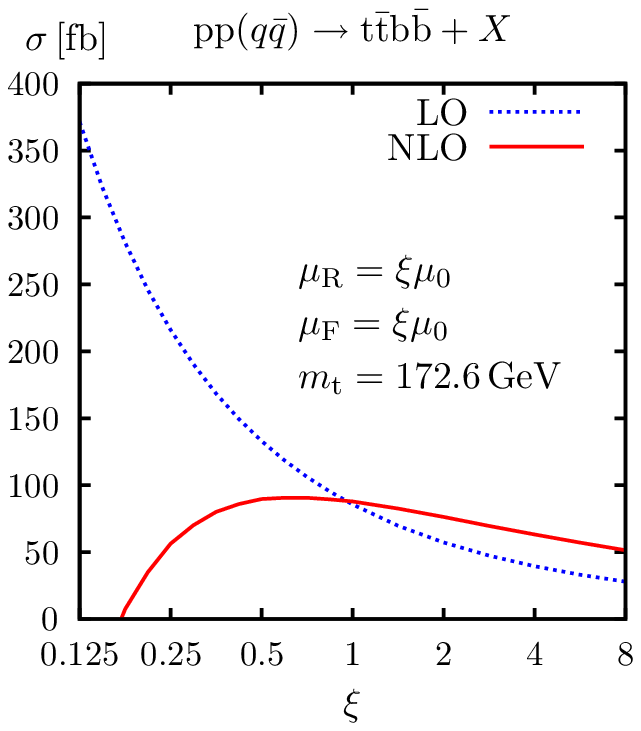}}
\put(80,0){\includegraphics[height=50mm,angle=0]{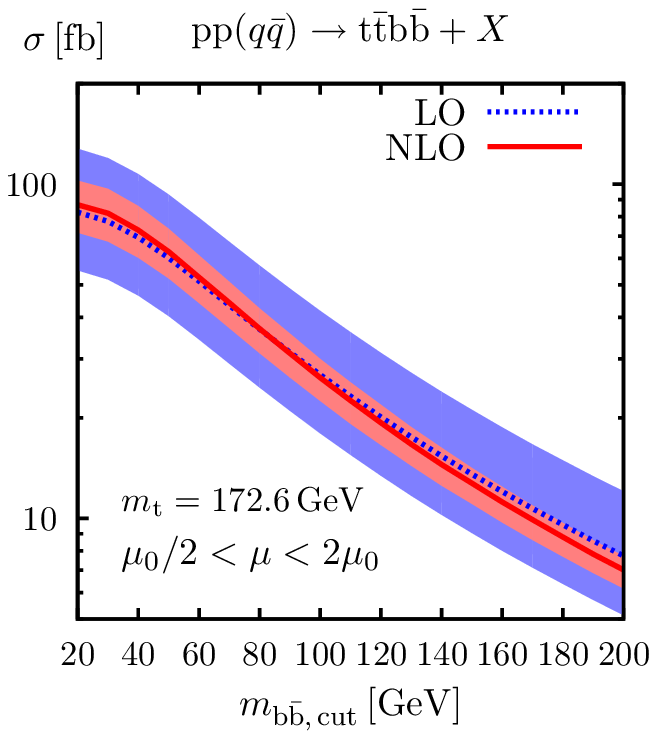}}
\end{picture}
\caption{Left: Comparison of the scale dependence for the LO (dashed)
and NLO (full) cross section
$pp(q\bar{q}) \to b\bar{b} t\bar{t}$ at LHC. Right: Effect of an invariant mass
cut on the b-pair. The scale variation is indicated by the two bands,
LO (blue), NLO (red) \cite{Bredenstein:2008zb}. }\label{qqbbtt}
\end{figure}
Note that the gluon induced subprocess  $gg\to b\bar{b} t\bar{t}$ has a more 
complicated tensor structure, nonetheless a full result for this important
Standard Model background can be expected in the near future.

\medskip

Another collaboration is currently pressing for an automated, Feynman diagrammatic
evaluation of multi-leg one-loop amplitudes. The {\tt GOLEM} collaboration,
where the acronym stands for General One-Loop  Evaluator of Matrix elements,
set up an automated reduction framework which takes special care of
numerical issues by providing alternatives for the evaluation
of one-loop integrals in critical phase space regions 
\cite{Binoth:1999sp,Binoth:2005ff,Binoth:2008uq}.
In this approach rank R N-point functions are reduced first to 
6-point tensor integrals and the latter are then expressed by Feynman 
parameter integrals  like
\begin{eqnarray}
I^{D,D+2}_{N=3,4}(j_1, \ldots ,j_r) &\sim&  \int_{0}^{1} 
\prod_{i=1}^{4} \, d z_i \, \delta(1-\sum_{l=1}^{4} z_l) 
\, \frac{z_{j_1} \ldots z_{j_r}}{ 
(-\frac{1}{2}\, z \cdot {\cal S}
\cdot z-i\delta)^{3-D/2}}\nonumber
\end{eqnarray}
Spurious numerical problems due to 
so called inverse Gram determinants are avoided by having the option
to evaluate the form factors in two different ways, either 
by numerical means or  by reducing the form factors numerically 
to a scalar integral basis. The method 
is designed to allow for an efficient isolation of IR and UV divergences
from the result and as such very well adapted for QCD calculations.   
The method can be also used to evaluate the rational part of an amplitude only 
\cite{Binoth:2006hk}.
All 6-point form factors are coded in a {\tt fortran} 95 code
{\tt golem95}, which is the first public library for such form factors \cite{Binoth:2008uq}.
The code {\tt golem95} relies on the evaluation of scalar integrals. In the present 
version only integrals with zero internal masses are implemented.
Another implementation of 6-point form factors has been presented
recently \cite{Diakonidis:2008ij,Diakonidis:2009yu}.

A full amplitude evaluation relies on diagrammatic input. The {\tt GOLEM}
collaboration uses public tools like {\tt QGRAF} \cite{Nogueira:1991ex} 
and {\tt FeynArts} \cite{Hahn:1998yk} for this step. 
The colour algebra and the helicity management is done as outlined
above. Subsequently two independent strategies are followed.
Firstly, the diagrammatic input is converted to a form factor representation
\begin{eqnarray}
\mathcal{A}^{\{\lambda\}} \to C_{box}^{ijk} I_4^{D+2,D+4}(i,j,k) 
+ C_{tri}^{ijk} I_3^{D,D+2}(i,j,k) + \dots \quad ,
\end{eqnarray}
exported to a {\tt fortran} code 
and then linked to the form factor library {\tt golem95}.
Secondly a fully symbolic representation of the scalar integral coefficients
is generated using {\tt FORM} \cite{Vermaseren:2000nd} and {\tt MAPLE}
\begin{eqnarray}
\mathcal{A}^{\{\lambda\}} \to C_{box} I_4^{D=6} 
                                                      + C_{tri} I_3^{D=4-2\epsilon}
                                                      + C_{bub} I_2^{D=4-2\epsilon} 
                                                      + C_{tad} I_1^{D=4-2\epsilon} + \cal{R}
\end{eqnarray}
Using these implementations
several computations of relevance for the LHC have been performed 
\cite{Binoth:2003xk,Binoth:2006mf,Binoth:2006ym,Andersen:2007mp}. The viability of the approach for $2\to 4$ processes 
was shown by the evaluation of the 6-photon 
amplitude mediated by a massless electron loop \cite{Binoth:2007ca}.
This amplitude has been used as testing ground for various methods
and perfect agreement has been obtained in all cases 
\cite{Nagy:2006xy,Ossola:2007bb,Bernicot:2007hs,Gong:2008ww}.

\medskip

\begin{figure}
\unitlength=1mm
\begin{picture}(100,26)
\put(5, 0){\includegraphics[height=25mm,angle=0]{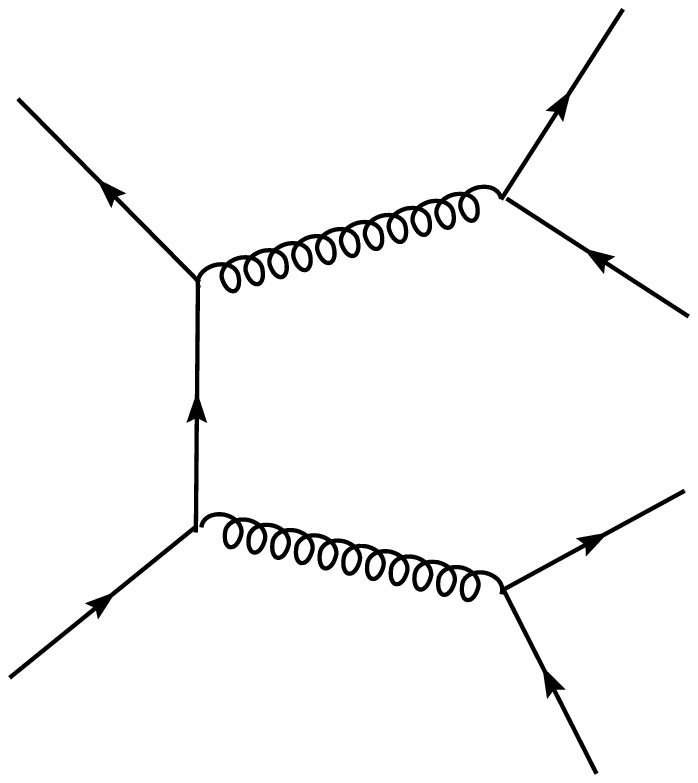}}
\put(40,0){\includegraphics[height=25mm,angle=0]{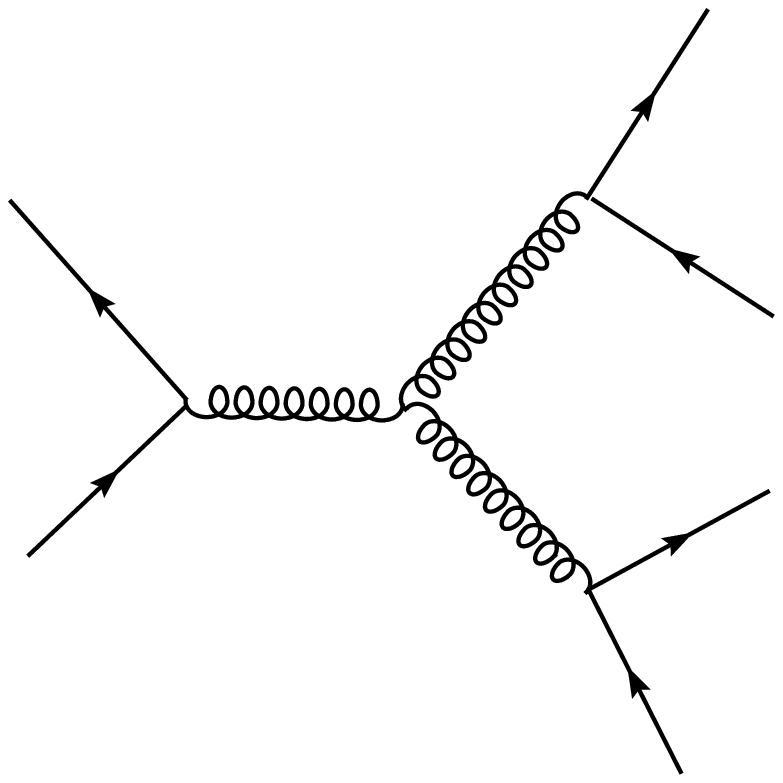}}
\put(80,0){\includegraphics[height=25mm,angle=0]{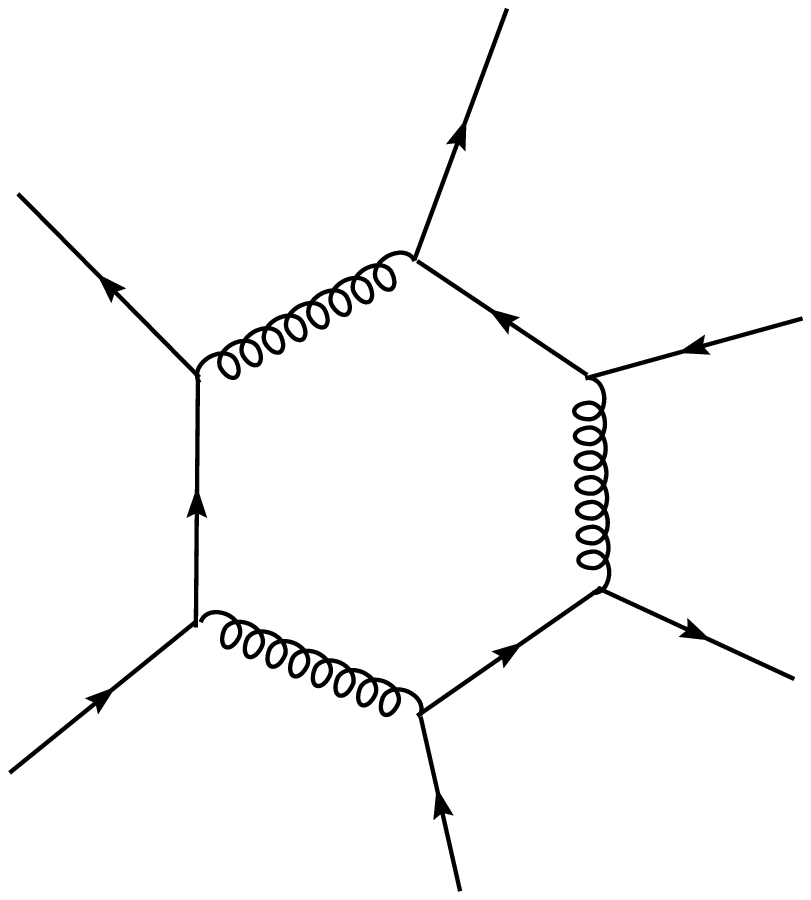}}
\put(115,0){\includegraphics[height=25mm,angle=0]{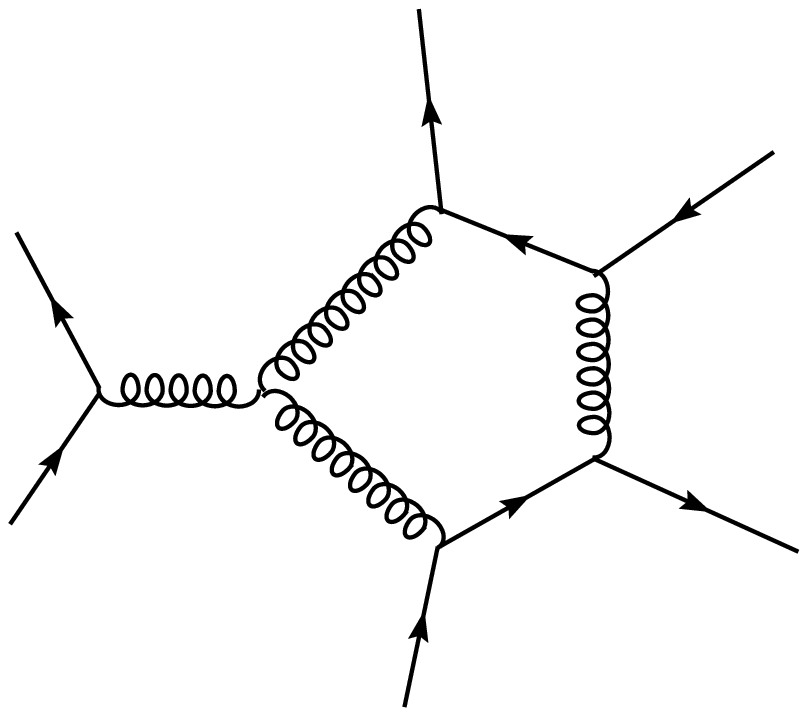}}
\end{picture}
\caption{LO and NLO topologies for the process $q\bar{q}\to b\bar{b} b\bar{b}$. }\label{qqbbbb_topos}
\end{figure}

\begin{figure}
\includegraphics[height=70mm]{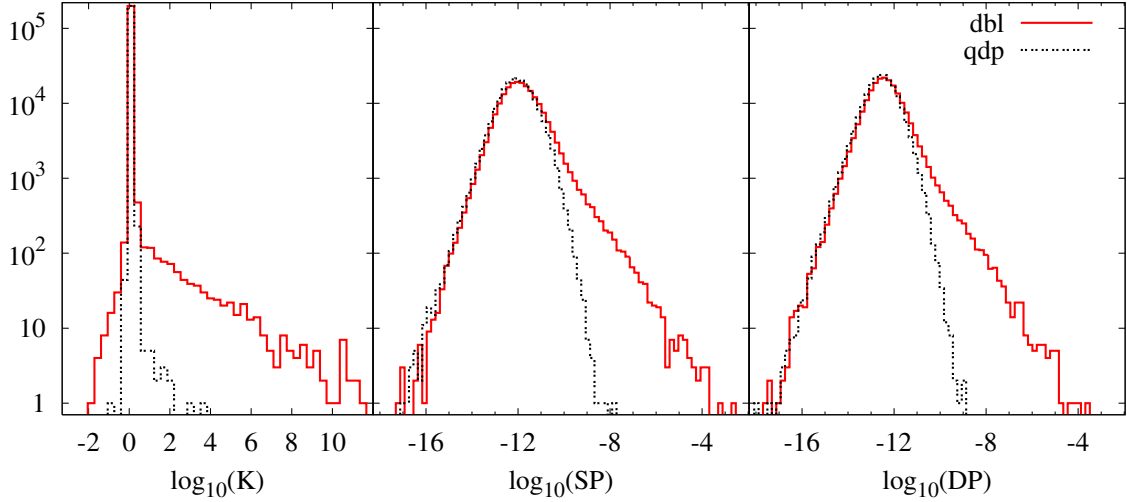}
\caption{Comparison of double (full) vs. quadruple precsion (dashed) evaluation
of the local K-factor function and the single/double pole zero coefficient
as defined in the text \cite{Reiter:2009kb}.}\label{reiter_numper}
\end{figure}

\begin{figure}
\unitlength=1mm
\begin{picture}(100,60)
\put(25,0){\includegraphics[height=60mm]{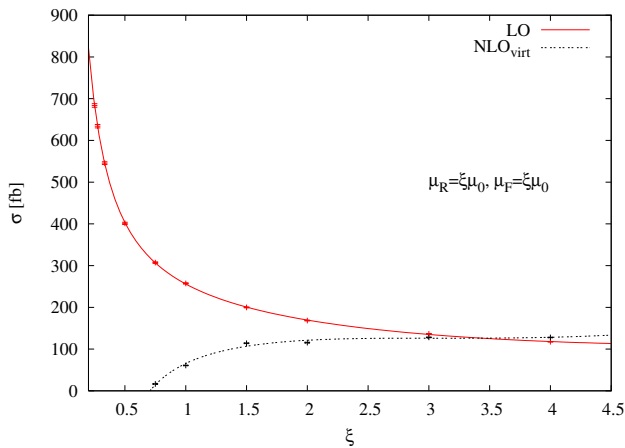}}
\end{picture}
\caption{Renormalisation scale dependence of the LO cross section (full red)
and including the UV/IR subtracted virtual correction term (dashed black) \cite{Reiter:2009kb}.}
\label{reiter_scaledep}
\end{figure}

Currently the NLO corrections for the process $pp\to b\bar{b} b\bar{b}$ are 
under construction using the developed computational tools. 
Because the 4b-quark background needs to be 
known as precisely 
as possible in the context of two Higgs doublet models, this process 
was included in the Les Houches ``experimentalist wish-list'' \cite{Bern:2008ef}.
The amplitude consists of two different initial states 
$q\bar{q}\to b\bar{b} b\bar{b}$ and $gg \to b\bar{b} b\bar{b}$.
The first can be represented by about 250 Feynman diagrams among which 
25 pentagon and 8 hexagon diagrams can be found, 
see Fig.~\ref{qqbbbb_topos} for a selection of LO and NLO diagrams.    
Note that only the pentagon and hexagon diagrams are computationally
challenging. The other topologies are relatively simple and do not
present a problem apart from book keeping which is not an issue
in automated approaches. The total number of diagrams is thus
not a good measure of the complexity of a calculation. 
All one-loop Feynman diagrams have been evaluated in two completely independent ways,
by symbolic reduction to scalar integrals and by using the form factor
decomposition in combination with the {\tt golem95} library as discussed above.
The virtual ${\cal O}(\alpha_s)$ correction is a interference term between  
the LO tree and NLO loop amplitude. After UV renormalisation, IR divergences
in form of $1/\epsilon$ single and double poles remain. The IR structure
of one-loop amplitudes is well-known and  a finite function can be obtained
after adding an adequate IR compensation term, as defined for example 
by the Catani-Seymour insertion operator 
$\langle {\cal A}_{LO}| {\bf I}(\epsilon) | {\cal A}_{LO}\rangle$ \cite{Catani:1996vz}.
In this way one can define a finite local K-factor function
\begin{equation}\label{kfac}
K = \frac{| {\cal A}_{LO} |^2 + 
2 \,\textrm{Re}( {\cal A}_{LO}^\dagger {\cal A}_{NLO, virt})
- \textrm{UV/IR subtractions}}{ | {\cal A}_{LO} |^2 } \quad .
\end{equation}   
To investigate the numerical performance of the approach, 200.000 random phase space points
have been evaluated using double and quadruple precision, see Fig.~\ref{reiter_numper}.
In the figure the numerical precision of the evaluation of the local K-factor
and the cancellation of the IR single/double pole are compared.
The result indicates that the size of the finite K-factor is a good indication
of the numerical accuracy of the evaluation. Large K-factors are 
mostly induced through numerical round-up errors. 
The cancellation of single and double poles, which is used by
other collaborations, can also be applied to decide about the
quality of the evaluation but seems to be less indicative. 
In Fig.~\ref{reiter_scaledep}
the improvement of the scale dependence is shown, if the
subtracted virtual correction term, as defined in 
Eq.~(\ref{kfac}), is included. The cross sections have been evaluated using the
experimental cuts $\eta < |2.5|$, $\Delta R > 0.4$, $p_T>25$ GeV
and the NLO pdf set CTEQ6.5.
A further improvement of the behaviour
is to be expected, because the logarithmic 
factorisation scale dependence is not yet compensated. 
To obtain the full NLO result one has to add the dipole subtracted NLO real emission 
corrections. 
For the full process, $pp\to b\bar{b}b \bar{b}$, the gluon induced
sub-process also has to be included. These parts of the calculation are in progress.  

Note that the evaluation strategy of the virtual NLO corrections
differs from the conventional approach of integrating a NLO cross section
directly. In the given example the LO cross section has been evaluated
first to produce an unweighted event sample. The subtracted 
virtual correction term defines a finite K-factor function which
is simply used to reweight these LO events. In this way
interference effects between adaptive Monte-Carlo integration 
and numerical round-up errors are completely avoided.
For a more detailed discussion see \cite{Binoth:2008gx}.
A similar strategy has also been applied in \cite{Lazopoulos:2007ix,Binoth:2008kt}.

\begin{figure}
\unitlength=1mm
\begin{picture}(100,60)
\put( 0,0){\includegraphics[height=6cm]{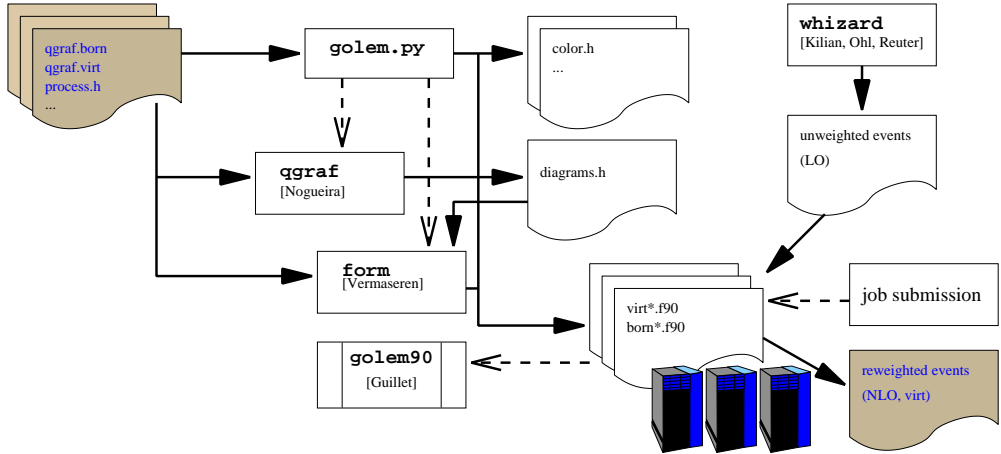}}
\end{picture}
\caption{Flow chart of the {\tt GOLEM} implementation \cite{Reiter:2009kb}.}\label{golemflow}
\end{figure}

The goal of the {\tt GOLEM} collaboration is to provide an
automated framework for one-loop amplitude computations.
The {\tt GOLEM} implementation is sketched in Fig.~\ref{golemflow}.
The input files which contain process and model dependent information
are processed with
a {\tt python} script which controls {\tt QGRAF} and {\tt FORM} codes.
The diagrammatic output, including UV and IR subtraction, is defined in terms 
of form factors and written to {\tt fortran} 95 code.
The latter is linked to the form factor library {\tt golem95}.
At the moment the code can be compiled in double or quadruple precision
but in the future a more dynamical solution will be adopted. 
The resulting computer program provides a function to
reweight LO event samples stemming from tree level event generators.
For the shown results  {\tt Whizard} \cite{Kilian:2007gr} has been used.

\subsection{Is there a preferred method?}

We have seen that both the unitarity based and also the
Feynman diagrammatic method are sufficiently well developed
meanwhile to allow for cross section calculations with more
than five external legs. While the unitarity based methods
were initially used to obtain compact analytical expressions for 
helicity amplitudes, the quest for automation
lead recently to various numerical implementations.
Higher precision libraries play a prominent role
in the related computer programs which shows that 
any given method seems to be doomed to deal with exceptional 
phase space regions in one way or the other.
Higher precision libraries seem to be a natural cure
for numerical problems. As the number of problematic 
phase space points is typically  only a few
percent of all points, increased evaluation times 
are not a severe issue\footnote{As discussed during this
conference a hardware implementation of quadruple precision
in processors is not yet in sight although technical
standards have been already defined for such a step.
The particle physics community would surely
be grateful for such a development. Unfortunately
fundamental science issues do not play the role they deserve
in business plans of computer chip producing companies,
although most technological developments are essentially
based on progress in fundamental science!}. 

\begin{figure}
\begin{picture}(120,60)
\put(30,0){\includegraphics[height=60mm,angle=0]{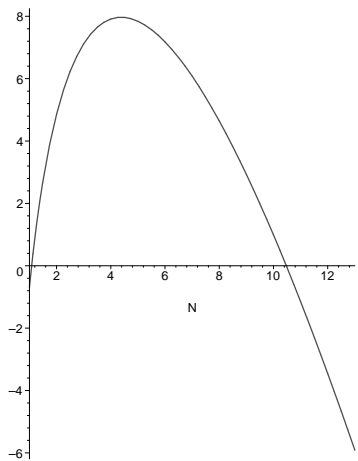}}
\end{picture}
\caption{The logarithmic ratio of the asymptotic behaviour, $\log( N^9/(\Gamma(N) \, 2^N) )$,
is plotted versus $N$.}\label{asy_ratio}
\end{figure}

Different methods will soon provide cross section predictions
for partonic $2\to 4$ processes relevant for LHC phenomenology.
As was explained above the scaling properties of one-loop 
N-point amplitude computations seem to indicate that the unitarity
based methods are clearly preferable. The factorial growth
in diagrammatic computations  has to be compared to
polynomial algorithms now.
For example, the evaluation of a colour ordered multi-gluon helicity amplitudes
scales like $N^9$ whereas a Feynman diagrammatic calculation
would involve about $2^N$ leg ordered diagrams with 
roughly $\Gamma(N)=(N-1)!$ form factor evaluations each.
However, let us ask the question, whether the asymptotic behaviour
gives a guideline which method is preferable for phenomenological 
applications.
The logarithmic ratio of the asymptotic behaviour, $\log( N^9/(\Gamma(N) \, 2^N) )$   
is plotted in Fig.~\ref{asy_ratio}.
The figure indicates that for phenomenological relevant multiplicities,
say $N\leq 8$, the asymptotic scaling behaviour is not a good measure
for the question which method is preferable. 
Furthermore the true complexity of a method should be measured
with full processes, taking into account all helicity amplitudes,
relevant colour structures and sub-processes.
It is important to note in this respect that
especially for background cross sections 
severe experimental cuts will be applied. Experience shows
that any  approximation or assumption might be invalidated in that way.
For the full task no ${\cal P}$-algorithm exists.

For $N\leq 8$, implementation issues, like efficient caching
and the re-use of recursive information can still be 
improved in Feynman diagram calculations. 
Together with the increasing computer power\footnote{We learnt at this
conference that the ``multi-core'' computer era will
change to the ``many-core'' era were the number of idle and available 
processors will increase enormously.} it can be expected
that both methods will deliver what they promise, namely
cross section predictions  for LHC phenomenology at the next-to-leading
order level.

\medskip

At the end of this section I would like to add the comment that fully numerical approaches
for Feynman diagram calculations have also been investigated
by several groups 
\cite{Ferroglia:2002mz,Binoth:2002xh,Kurihara:2005ja,Nagy:2006xy,Anastasiou:2007qb,Gong:2008ww}. Given the huge world-wide computing resources
any reliable integration method of multi-parameter integrals
containing threshold singularities could lead to a 
solution of the one-loop problem without using any
algebraic overhead. At the moment these methods are
not competitive with the discussed approaches but
there is certainly room for large improvements. 

\section{Conclusion and Outlook}

In this talk I have argued that the LHC needs and
deserves an effort to predict prominent signal and background
processes at the next-to-leading order level in QCD.
Especially multi-particle amplitudes  have a high
power of $\alpha_s(\mu)$ which induce large scale variations.
Absolute rates thus can not be predicted reliably with 
leading order Monte-Carlo 
tools and eventually this will hamper the understanding of LHC
data and the discovery of new physics. Many relevant 
Standard Model processes with three final 
state particles are meanwhile available in the literature
beyond the leading order.
Unfortunately the results are not always available as a public
code which would be most beneficial for the experimental community. 
The theoretical community is at the moment enormously active
to go beyond the given level of complexity and  
NLO predictions for processes like $pp\to jjjj$, $pp\to Wjjj$, 
$pp\to WWjj$, $pp\to b\bar{b} t\bar{t}$, $pp\to jj t\bar{t}$, 
$pp\to b\bar{b} b\bar{b}$, which are highly relevant for 
various Higgs and BSM search channels, 
are under construction. Many different methods have been 
developed and are applied for the virtual corrections.
Unitarity based methods look very promising in this respect but
for the required level of complexity both, the 
Feynman diagram and the unitarity based approaches, will 
provide valuable results for phenomenology. 
The avenue to fully numerical approaches to one-loop 
amplitudes is still not fully explored, but given the 
ever growing computer power this might be a promising
direction avoiding any kind of algebraic manipulations
of initial expressions.

Most groups move at the moment towards automated approaches
which will provide platforms to do many computations
with the same set-up. To cure the problem of numerical instabilities 
multi-precision libraries seem to be the accepted way out now.
It would be highly appreciable, if the community could agree on a 
standardisation of in- and output 
formats for NLO codes and would make computer programs
publicly available. The structure of one-loop computations
is indeed very modular and a minimal agreement on 
colour and helicity management and on passing IR subtraction
terms, which are basically universal anyhow, could result
in transportable modules for virtual corrections.  
This would allow to use computations of different groups interchangeably
by combining them with public tree level matrix element generators
which should of course also contain the functionality of IR subtractions.
In a next step the combination with parton showers could be
obtained. Here the inclusion of NLO precision is also well understood, 
but again a focus on modularity and transportability would help
to use synergies between different groups. 

The main conclusion of this talk is that 
the developments of the last few years
are spectacular and point 
towards Monte Carlo tools at full 
next-to-leading order QCD level, at least as
long the number of final state particles is not 
excessively high. The dominant and most relevant
processes for Higgs and new physisc searches
will certainly be available in form of flexible
and reliable public computer programs and 
the analysis of LHC data beyond the leading order
in $\alpha_s$ will eventually become the standard.

\section*{Acknowledgements}   
I would like to thank the organisers of the ACAT 2008
workshop for the opportunity to present my view
on the progress of the field during the fantastic meeting in Erice, Italy. 
I also would like to thank my colleagues of the 
{\tt GOLEM} collaboration, A.~Guffanti, J.~Ph.~Guillet, G.~Heinrich, S.~Karg,
N.~Kauer, T.~Reiter, J.~Reuter and G.~Sanguinetti for many stimulating 
discussions and their cooperation, which helped to shape this 
contribution in a substantial way.


\begin{thebibliography}{150}

%\cite{Djouadi:2005gi}
\bibitem{Djouadi:2005gi}
  A.~Djouadi,
  %``The anatomy of electro-weak symmetry breaking. I: The Higgs boson in  the
  %standard model,''
  Phys.\ Rept.\  {\bf 457} (2008) 1
  [arXiv:hep-ph/0503172].
  %%CITATION = PRPLC,457,1;%%

%\cite{:2008uu}
\bibitem{:2008uu}
  N.~E.~Adam {\it et al.},
  %``Higgs Working Group Summary Report,''
  arXiv:0803.1154 [hep-ph].
 

%\cite{Martin:1997ns}
\bibitem{Martin:1997ns}
  S.~P.~Martin,
  %``A Supersymmetry Primer,''
  arXiv:hep-ph/9709356.
  %%CITATION = HEP-PH/9709356;%%


%\cite{Djouadi:2005gj}
\bibitem{Djouadi:2005gj}
  A.~Djouadi,
  %``The anatomy of electro-weak symmetry breaking. II: The Higgs bosons in  the
  %minimal supersymmetric model,''
  Phys.\ Rept.\  {\bf 459} (2008) 1
  [arXiv:hep-ph/0503173].
  %%CITATION = PRPLC,459,1;%%




  
%CIT%\cite{Allanach:2006fy}
\bibitem{Allanach:2006fy}
  B.~C.~Allanach {\it et al.},
  %``Les Houches 'Physics at TeV colliders 2005' Beyond the standard model
  %working group: Summary report,''
  arXiv:hep-ph/0602198.
  %%CITATION = HEP-PH/0602198;%%
%???ATION = ARXIV:0803.1154;%%


%\cite{Sundrum:2005jf}
\bibitem{Sundrum:2005jf}
  R.~Sundrum, (TASI 2004), 
  %``To the fifth dimension and back. (TASI 2004),''
  arXiv:hep-th/0508134.
  %%CITATION = HEP-TH/0508134;%%

%\cite{Csaki:2005vy}
\bibitem{Csaki:2005vy}
  C.~Csaki, J.~Hubisz and P.~Meade, (TASI 2005)
  %``Electroweak symmetry breaking from extra dimensions,''
  arXiv:hep-ph/0510275.
  %%CITATION = HEP-PH/0510275;%%

%\cite{Kribs:2006mq}
\bibitem{Kribs:2006mq}
  G.~D.~Kribs, (TASI 2006)
  %``Phenomenology of extra dimensions,''
  arXiv:hep-ph/0605325.
  %%CITATION = HEP-PH/0605325;%%

%\cite{Schmaltz:2005ky}
\bibitem{Schmaltz:2005ky}
  M.~Schmaltz and D.~Tucker-Smith,
  %``Little Higgs Review,''
  Ann.\ Rev.\ Nucl.\ Part.\ Sci.\  {\bf 55} (2005) 229
  [arXiv:hep-ph/0502182].
  %%CITATION = ARNUA,55,229;%%


%\cite{atlas:old}
\bibitem{atlas:old}
K. Cranmer, B. Mellado, W. Quayle, S.L. Wu, ``Statistical Methods to Assess
the Combined Sensitivity of the ATLAS Detector to the Higgs Boson in the
Standard Model'', ATL-PHYS-2005-034.

%\cite{cms:old}
\bibitem{cms:old}
CMS Collaboration, ``CMS Physics Technical Design Report Volume II:
Physics Performance'',
CERN/LHCC/2006-021, CMS TDR 8.2.


%\cite{Jakobs:2007zza}
\bibitem{Jakobs:2007zza}
  K.~Jakobs  [ATLAS Collaboration],
  %``Prospects for Higgs boson searches at the LHC,''
  Int.\ J.\ Mod.\ Phys.\  A {\bf 23}, 5093 (2008).
  %%CITATION = IMPAE,A23,5093;%%

%\cite{Buescher:2005re}
\bibitem{Buescher:2005re}
  V.~Buescher and K.~Jakobs,
  %``Higgs boson searches at hadron colliders,''
  Int.\ J.\ Mod.\ Phys.\  A {\bf 20}, 2523 (2005)
  [arXiv:hep-ph/0504099].
  %%CITATION = IMPAE,A20,2523;%%

%\cite{Aad:2009wy}
\bibitem{Aad:2009wy}
  G.~Aad {\it et al.}  [The ATLAS Collaboration],
  %``Expected Performance of the ATLAS Experiment - Detector, Trigger and
  %Physics,''
  arXiv:0901.0512.
  %%CITATION = ARXIV:0901.0512;%%

%\cite{Binoth:1999qq}
\bibitem{Binoth:1999qq}
  T.~Binoth, J.~P.~Guillet, E.~Pilon and M.~Werlen,
  %``A full next to leading order study of direct photon pair production in
  %hadronic collisions,''
  Eur.\ Phys.\ J.\  C {\bf 16} (2000) 311
  [arXiv:hep-ph/9911340].
  %%CITATION = EPHJA,C16,311;%%

%\cite{Bern:2002jx}
\bibitem{Bern:2002jx}
  Z.~Bern, L.~J.~Dixon and C.~Schmidt,
  %``Isolating a light Higgs boson from the di-photon background at the LHC,''
  Phys.\ Rev.\  D {\bf 66} (2002) 074018
  [arXiv:hep-ph/0206194].
  %%CITATION = PHRVA,D66,074018;%%

%\cite{Kauer:2000hi}
\bibitem{Kauer:2000hi}
  N.~Kauer, T.~Plehn, D.~L.~Rainwater and D.~Zeppenfeld,
  %``H --> W W as the discovery mode for a light Higgs boson,''
  Phys.\ Lett.\  B {\bf 503}, 113 (2001)
  [arXiv:hep-ph/0012351].
  %%CITATION = PHLTA,B503,113;%%




%\cite{Mangano:2002ea}
\bibitem{Mangano:2002ea}
  M.~L.~Mangano, M.~Moretti, F.~Piccinini, R.~Pittau and A.~D.~Polosa,
  %``ALPGEN, a generator for hard multiparton processes in hadronic
  %collisions,''
  JHEP {\bf 0307} (2003) 001
  [arXiv:hep-ph/0206293].
  %%CITATION = JHEPA,0307,001;%%

%\cite{Pukhov:1999gg}
\bibitem{Pukhov:1999gg}
  A.~Pukhov {\it et al.},
  %``CompHEP: A package for evaluation of Feynman diagrams and integration  over
  %multi-particle phase space. User's manual for version 33,''
  arXiv:hep-ph/9908288.
  %%CITATION = HEP-PH/9908288;%%

%\cite{Krauss:2001iv}
\bibitem{Krauss:2001iv}
  F.~Krauss, R.~Kuhn and G.~Soff,
  %``AMEGIC++ 1.0: A matrix element generator in C++,''
  JHEP {\bf 0202} (2002) 044
  [arXiv:hep-ph/0109036].
  %%CITATION = JHEPA,0202,044;%%


%\cite{Gleisberg:2008fv}
\bibitem{Gleisberg:2008fv}
  T.~Gleisberg and S.~Hoche,
  %``Comix, a new matrix element generator,''
  JHEP {\bf 0812} (2008) 039
  [arXiv:0808.3674 [hep-ph]].
  %%CITATION = JHEPA,0812,039;%%

%\cite{Hahn:1998yk}
\bibitem{Hahn:1998yk}
  T.~Hahn and M.~Perez-Victoria,
  %``Automatized one-loop calculations in four and D dimensions,''
  Comput.\ Phys.\ Commun.\  {\bf 118}, 153 (1999)
  [arXiv:hep-ph/9807565].
  %%CITATION = CPHCB,118,153;%%

%\cite{Belanger:2003sd}
\bibitem{Belanger:2003sd}
  G.~Belanger, F.~Boudjema, J.~Fujimoto, T.~Ishikawa, T.~Kaneko, K.~Kato and Y.~Shimizu,
  %``Automatic calculations in high energy physics and Grace at one-loop,''
  Phys.\ Rept.\  {\bf 430}, 117 (2006)
  [arXiv:hep-ph/0308080].
  %%CITATION = PRPLC,430,117;%%

%\cite{Kanaki:2000ey}
\bibitem{Kanaki:2000ey}
  A.~Kanaki and C.~G.~Papadopoulos,
  %``HELAC: A package to compute electroweak helicity amplitudes,''
  Comput.\ Phys.\ Commun.\  {\bf 132}, 306 (2000)
  [arXiv:hep-ph/0002082].
  %%CITATION = CPHCB,132,306;%%

%\cite{Papadopoulos:2006mh}
\bibitem{Papadopoulos:2006mh}
  C.~G.~Papadopoulos and M.~Worek,
  %``HELAC: A Monte Carlo generator for multi-jet processes,''
  arXiv:hep-ph/0606320.
  %%CITATION = HEP-PH/0606320;%%

%\cite{Cafarella:2007pc}
\bibitem{Cafarella:2007pc}
  A.~Cafarella, C.~G.~Papadopoulos and M.~Worek,
  %``Helac-Phegas: a generator for all parton level processes,''
  arXiv:0710.2427 [hep-ph].
  %%CITATION = ARXIV:0710.2427;%%


%\cite{Maltoni:2002qb}
\bibitem{Maltoni:2002qb}
  F.~Maltoni and T.~Stelzer,
  %``MadEvent: Automatic event generation with MadGraph,''
  JHEP {\bf 0302}, 027 (2003)
  [arXiv:hep-ph/0208156].
  %%CITATION = JHEPA,0302,027;%%

%\cite{Stelzer:1994ta}
\bibitem{Stelzer:1994ta}
  T.~Stelzer and W.~F.~Long,
  %``Automatic generation of tree level helicity amplitudes,''
  Comput.\ Phys.\ Commun.\  {\bf 81}, 357 (1994)
  [arXiv:hep-ph/9401258].
  %%CITATION = CPHCB,81,357;%%

%\cite{Kilian:2007gr}
\bibitem{Kilian:2007gr}
  W.~Kilian, T.~Ohl and J.~Reuter,
  %``WHIZARD: Simulating Multi-Particle Processes at LHC and ILC,''
  arXiv:0708.4233 [hep-ph].
  %%CITATION = ARXIV:0708.4233;%%

%\cite{Sjostrand:2006za}
\bibitem{Sjostrand:2006za}
  T.~Sjostrand, S.~Mrenna and P.~Skands,
  %``PYTHIA 6.4 physics and manual,''
  JHEP {\bf 0605}, 026 (2006)
  [arXiv:hep-ph/0603175].
  %%CITATION = JHEPA,0605,026;%%

%\cite{Bahr:2008pv}
\bibitem{Bahr:2008pv}
  M.~Bahr {\it et al.},
  %``Herwig++ Physics and Manual,''
  Eur.\ Phys.\ J.\  C {\bf 58}, 639 (2008)
  [arXiv:0803.0883 [hep-ph]].
  %%CITATION = EPHJA,C58,639;%%

%\cite{Gleisberg:2008ta}
\bibitem{Gleisberg:2008ta}
  T.~Gleisberg, S.~Hoche, F.~Krauss, M.~Schonherr, S.~Schumann, F.~Siegert and J.~Winter,
  %``Event generation with SHERPA 1.1,''
  JHEP {\bf 0902}, 007 (2009)
  [arXiv:0811.4622 [hep-ph]].
  %%CITATION = JHEPA,0902,007;%%

%\cite{Catani:1996vz}
\bibitem{Catani:1996vz}
  S.~Catani and M.~H.~Seymour,
  %``A general algorithm for calculating jet cross sections in NLO QCD,''
  Nucl.\ Phys.\  B {\bf 485}, 291 (1997)
  [Erratum-ibid.\  B {\bf 510}, 503 (1998)]
  [arXiv:hep-ph/9605323].
  %%CITATION = NUPHA,B485,291;%%

%\cite{Catani:2002hc}
\bibitem{Catani:2002hc}
  S.~Catani, S.~Dittmaier, M.~H.~Seymour and Z.~Trocsanyi,
  %``The dipole formalism for next-to-leading order QCD calculations with
  %massive partons,''
  Nucl.\ Phys.\  B {\bf 627}, 189 (2002)
  [arXiv:hep-ph/0201036].
  %%CITATION = NUPHA,B627,189;%%

%\cite{Seymour:2008mu}
\bibitem{Seymour:2008mu}
  M.~H.~Seymour and C.~Tevlin,
  %``TeVJet: A general framework for the calculation of jet observables in NLO
  %QCD,''
  arXiv:0803.2231 [hep-ph].
  %%CITATION = ARXIV:0803.2231;%%

%\cite{Gleisberg:2007md}
\bibitem{Gleisberg:2007md}
  T.~Gleisberg and F.~Krauss,
  %``Automating dipole subtraction for QCD NLO calculations,''
  Eur.\ Phys.\ J.\  C {\bf 53}, 501 (2008)
  [arXiv:0709.2881 [hep-ph]].
  %%CITATION = EPHJA,C53,501;%%

%\cite{Hasegawa:2008ae}
\bibitem{Hasegawa:2008ae}
  K.~Hasegawa, S.~Moch and P.~Uwer,
  %``Automating dipole subtraction,''
  Nucl.\ Phys.\ Proc.\ Suppl.\  {\bf 183}, 268 (2008)
  [arXiv:0807.3701 [hep-ph]].
  %%CITATION = NUPHZ,183,268;%%

%\cite{Frederix:2008hu}
\bibitem{Frederix:2008hu}
  R.~Frederix, T.~Gehrmann and N.~Greiner,
  %``Automation of the Dipole Subtraction Method in MadGraph/MadEvent,''
  JHEP {\bf 0809}, 122 (2008)
  [arXiv:0808.2128 [hep-ph]].
  %%CITATION = JHEPA,0809,122;%%


%\cite{Campbell:2000bg}
\bibitem{Campbell:2000bg}
  J.~M.~Campbell and R.~K.~Ellis,
  %``Radiative corrections to Z b anti-b production,''
  Phys.\ Rev.\  D {\bf 62} (2000) 114012
  [arXiv:hep-ph/0006304].
  %%CITATION = PHRVA,D62,114012;%%

\bibitem{mcfm_page}
http://mcfm.fnal.gov/

\bibitem{phox_page}
%\begin{verbatim}
http://wwwlapp.in2p3.fr/lapth/PHOX FAMILY/main.html
%\end{verbatim}

%\cite{Arnold:2008rz}
\bibitem{Arnold:2008rz}
  K.~Arnold {\it et al.},
  %``VBFNLO: A parton level Monte Carlo for processes with electroweak bosons,''
  arXiv:0811.4559 [hep-ph].
  %%CITATION = ARXIV:0811.4559;%%

%\cite{Frixione:2008ym}
\bibitem{Frixione:2008ym}
  S.~Frixione and B.~R.~Webber,
  %``The MC@NLO 3.4 Event Generator,''
  arXiv:0812.0770 [hep-ph].
  %%CITATION = ARXIV:0812.0770;%%

%\cite{Frixione:2007vw}
\bibitem{Frixione:2007vw}
  S.~Frixione, P.~Nason and C.~Oleari,
  %``Matching NLO QCD computations with Parton Shower simulations: the POWHEG
  %method,''
  JHEP {\bf 0711}, 070 (2007)
  [arXiv:0709.2092 [hep-ph]].
  %%CITATION = JHEPA,0711,070;%%


%\cite{Fujimoto:2008zz}
\bibitem{Fujimoto:2008zz}
  J.~Fujimoto and Y.~Kurihara,
  %``GRACE-NLO for the LHC,''
  Nucl.\ Phys.\ Proc.\ Suppl.\  {\bf 183}, 143 (2008).
  %%CITATION = NUPHZ,183,143;%%

%\cite{Nagy:2005aa}
\bibitem{Nagy:2005aa}
  Z.~Nagy and D.~E.~Soper,
  %``Matching parton showers to NLO computations,''
  JHEP {\bf 0510} (2005) 024
  [arXiv:hep-ph/0503053].
  %%CITATION = JHEPA,0510,024;%%



%\cite{Lavesson:2008ah}
\bibitem{Lavesson:2008ah}
  N.~Lavesson and L.~Lonnblad,
  %``Extending CKKW-merging to One-Loop Matrix Elements,''
  JHEP {\bf 0812} (2008) 070
  [arXiv:0811.2912 [hep-ph]].
  %%CITATION = JHEPA,0812,070;%%

%\cite{Andonov:2003xe}
\bibitem{Andonov:2003xe}
  A.~Andonov, D.~Bardin, S.~Bondarenko, P.~Christova, L.~Kalinovskaya, G.~Nanava and G.~Passarino,
  %``SANC press release,''
  Nucl.\ Instrum.\ Meth.\  A {\bf 502} (2003) 576.
  %%CITATION = NUIMA,A502,576;%%


%\cite{Andonov:2007zz}
\bibitem{Andonov:2007zz}
  A.~Andonov, A.~Arbuzov, S.~Bondarenko, P.~Christova, V.~Kolesnikov and R.~Sadykov,
  %``Implementation Of Nlo QCD Corrections Into The Framework Of Computer System
  %Sanc,''
  Phys.\ Part.\ Nucl.\ Lett.\  {\bf 4} (2007) 451.
  %%CITATION = 00438,4,451;%%

%\cite{Bardin:2007zz}
\bibitem{Bardin:2007zz}
  D.~Bardin {\it et al.},
  %``SANC: precision calculations for the SM processes,''
  PoS A {\bf CAT} (2007) 077.
  %%CITATION = POSCI,ACAT,077;%%

  %\cite{Binoth:2005ua}
\bibitem{Binoth:2005ua}
  T.~Binoth, M.~Ciccolini, N.~Kauer and M.~Kramer,
  %``Gluon-induced W W background to Higgs boson searches at the LHC,''
  JHEP {\bf 0503}, 065 (2005)
  [arXiv:hep-ph/0503094].
  %%CITATION = JHEPA,0503,065;%%

%\cite{Binoth:2006mf}
\bibitem{Binoth:2006mf}
  T.~Binoth, M.~Ciccolini, N.~Kauer and M.~Kramer,
  %``Gluon-induced W-boson pair production at the LHC,''
  JHEP {\bf 0612}, 046 (2006)
  [arXiv:hep-ph/0611170].
  %%CITATION = JHEPA,0612,046;%%

%\cite{Binoth:2008pr}
\bibitem{Binoth:2008pr}
  T.~Binoth, N.~Kauer and P.~Mertsch,
  %``Gluon-induced QCD corrections to pp --> ZZ --> l anti-l l' anti-l',''
  arXiv:0807.0024 [hep-ph].
  %%CITATION = ARXIV:0807.0024;%%

  
%\cite{Nagy:2003tz}
\bibitem{Nagy:2003tz}
  Z.~Nagy,
  %``Next-to-leading order calculation of three-jet observables in hadron-hadron
  %collision,''
  Phys.\ Rev.\  D {\bf 68}, 094002 (2003)
  [arXiv:hep-ph/0307268].
  %%CITATION = PHRVA,D68,094002;%%

%\cite{deFlorian:1999tp}
\bibitem{deFlorian:1999tp}
  D.~de Florian and Z.~Kunszt,
  %``Two photons plus jet at LHC: The NNLO contribution from the g g  initiated
  %process,''
  Phys.\ Lett.\  B {\bf 460}, 184 (1999)
  [arXiv:hep-ph/9905283].
  %%CITATION = PHLTA,B460,184;%%


%\cite{DelDuca:2003uz}
\bibitem{DelDuca:2003uz}
  V.~Del Duca, F.~Maltoni, Z.~Nagy and Z.~Trocsanyi,
  %``QCD radiative corrections to prompt diphoton production in association
  %with a jet at hadron colliders,''
  JHEP {\bf 0304}, 059 (2003)
  [arXiv:hep-ph/0303012].
  %%CITATION = JHEPA,0304,059;%%


%\cite{Campbell:2006xx}
\bibitem{Campbell:2006xx}
  J.~M.~Campbell, R.~K.~Ellis and G.~Zanderighi,
  %``Next-to-leading order Higgs + 2 jet production via gluon fusion,''
  JHEP {\bf 0610}, 028 (2006)
  [arXiv:hep-ph/0608194].
  %%CITATION = JHEPA,0610,028;%%

%\cite{Han:1991ia}
\bibitem{Han:1991ia}
  T.~Han and S.~Willenbrock,
  %``QCD correction to the p p $\to$ W H and Z H total cross-sections,''
  Phys.\ Lett.\  B {\bf 273} (1991) 167.
  %%CITATION = PHLTA,B273,167;%%

%\cite{Figy:2003nv}
\bibitem{Figy:2003nv}
  T.~Figy, C.~Oleari and D.~Zeppenfeld,
  %``Next-to-leading order jet distributions for Higgs boson production via
  %weak-boson fusion,''
  Phys.\ Rev.\  D {\bf 68} (2003) 073005
  [arXiv:hep-ph/0306109].
  %%CITATION = PHRVA,D68,073005;%%

%\cite{Beenakker:2002nc}
\bibitem{Beenakker:2002nc}
  W.~Beenakker, S.~Dittmaier, M.~Kramer, B.~Plumper, M.~Spira and P.~M.~Zerwas,
  %``NLO QCD corrections to t anti-t H production in hadron collisions. ((U)),''
  Nucl.\ Phys.\  B {\bf 653} (2003) 151
  [arXiv:hep-ph/0211352].
  %%CITATION = NUPHA,B653,151;%%

%\cite{Dawson:2003hv}
\bibitem{Dawson:2003hv}
  S.~Dawson, C.~B.~Jackson, L.~H.~Orr, L.~Reina and D.~Wackeroth,
  %``Theoretical progress for the associated production of a Higgs boson  with
  %heavy quarks at hadron colliders,''
  Eur.\ Phys.\ J.\  C {\bf 33}, S451 (2004)
  [arXiv:hep-ph/0311216].
  %%CITATION = EPHJA,C33,S451;%%

%\cite{Plehn:2005nk}
\bibitem{Plehn:2005nk}
  T.~Plehn and M.~Rauch,
  %``The quartic Higgs coupling at hadron colliders,''
  Phys.\ Rev.\  D {\bf 72} (2005) 053008
  [arXiv:hep-ph/0507321].
  %%CITATION = PHRVA,D72,053008;%%


%\cite{Binoth:2006ym}
\bibitem{Binoth:2006ym}
  T.~Binoth, S.~Karg, N.~Kauer and R.~Ruckl,
  %``Multi-Higgs boson production in the standard model and beyond,''
  Phys.\ Rev.\  D {\bf 74} (2006) 113008
  [arXiv:hep-ph/0608057].
  %%CITATION = PHRVA,D74,113008;%%

%\cite{Jager:2006zc}
\bibitem{Jager:2006zc}
  B.~Jager, C.~Oleari and D.~Zeppenfeld,
  %``Next-to-leading order QCD corrections to W+ W- production via vector-boson
  %fusion,''
  JHEP {\bf 0607} (2006) 015
  [arXiv:hep-ph/0603177].
  %%CITATION = JHEPA,0607,015;%%

%\cite{Jager:2006cp}
\bibitem{Jager:2006cp}
  B.~Jager, C.~Oleari and D.~Zeppenfeld,
  %``Next-to-leading order QCD corrections to Z boson pair production via
  %vector-boson fusion,''
  Phys.\ Rev.\  D {\bf 73} (2006) 113006
  [arXiv:hep-ph/0604200].
  %%CITATION = PHRVA,D73,113006;%%


%\cite{Bozzi:2007ur}
\bibitem{Bozzi:2007ur}
  G.~Bozzi, B.~Jager, C.~Oleari and D.~Zeppenfeld,
  %``Next-to-leading order QCD corrections to W+Z and W-Z production via
  %vector-boson fusion,''
  Phys.\ Rev.\  D {\bf 75} (2007) 073004
  [arXiv:hep-ph/0701105].
  %%CITATION = PHRVA,D75,073004;%%

%\cite{Lazopoulos:2007ix}
\bibitem{Lazopoulos:2007ix}
  A.~Lazopoulos, K.~Melnikov and F.~Petriello,
  %``QCD corrections to tri-boson production,''
  Phys.\ Rev.\  D {\bf 76} (2007) 014001
  [arXiv:hep-ph/0703273].
  %%CITATION = PHRVA,D76,014001;%%

%\cite{Dittmaier:2007wz}
\bibitem{Dittmaier:2007wz}
  S.~Dittmaier, P.~Uwer and S.~Weinzierl,
  %``NLO QCD corrections to t anti-t + jet production at hadron colliders,''
  Phys.\ Rev.\ Lett.\  {\bf 98} (2007) 262002
  [arXiv:hep-ph/0703120].
  %%CITATION = PRLTA,98,262002;%%

%\cite{Dittmaier:2008uj}
\bibitem{Dittmaier:2008uj}
  S.~Dittmaier, P.~Uwer and S.~Weinzierl,
  %``Hadronic top-quark pair production in association with a hard jet at
  %next-to-leading order QCD: Phenomenological studies for the Tevatron and the
  %LHC,''
  arXiv:0810.0452 [hep-ph].
  %%CITATION = ARXIV:0810.0452;%%

%\cite{Dittmaier:2007th}
\bibitem{Dittmaier:2007th}
  S.~Dittmaier, S.~Kallweit and P.~Uwer,
  %``NLO QCD corrections to WW+jet production at hadron colliders,''
  arXiv:0710.1577 [hep-ph].
  %%CITATION = ARXIV:0710.1577;%%

%\cite{Campbell:2007ev}
\bibitem{Campbell:2007ev}
  J.~M.~Campbell, R.~Keith Ellis and G.~Zanderighi,
  %``Next-to-leading order predictions for $WW+1$ jet distributions at the
  %LHC,''
  JHEP {\bf 0712} (2007) 056
  [arXiv:0710.1832 [hep-ph]].
  %%CITATION = JHEPA,0712,056;%%

%\cite{Binoth:2008kt}
\bibitem{Binoth:2008kt}
  T.~Binoth, G.~Ossola, C.~G.~Papadopoulos and R.~Pittau,
  %``NLO QCD corrections to tri-boson production,''
  JHEP {\bf 0806}, 082 (2008)
  [arXiv:0804.0350 [hep-ph]].
  %%CITATION = JHEPA,0806,082;%%

%\cite{Hankele:2007sb}
\bibitem{Hankele:2007sb}
  V.~Hankele and D.~Zeppenfeld,
  %``QCD corrections to hadronic WWZ production with leptonic decays,''
  Phys.\ Lett.\  B {\bf 661} (2008) 103
  [arXiv:0712.3544 [hep-ph]].
  %%CITATION = PHLTA,B661,103;%%


%\cite{Lazopoulos:2007bv}
\bibitem{Lazopoulos:2007bv}
  A.~Lazopoulos, K.~Melnikov and F.~J.~Petriello,
  %``NLO QCD corrections to the production of t-tbar-Z in gluon fusion,''
  Phys.\ Rev.\  D {\bf 77}, 034021 (2008)
  [arXiv:0709.4044 [hep-ph]].
  %%CITATION = PHRVA,D77,034021;%%

%\cite{Lazopoulos:2008de}
\bibitem{Lazopoulos:2008de}
  A.~Lazopoulos, T.~McElmurry, K.~Melnikov and F.~Petriello,
  %``Next-to-leading order QCD corrections to $t \bar{t} Z$ production at the
  %LHC,''
  Phys.\ Lett.\  B {\bf 666} (2008) 62
  [arXiv:0804.2220 [hep-ph]].
  %%CITATION = PHLTA,B666,62;%%


%\cite{FebresCordero:2008ci}
\bibitem{FebresCordero:2008ci}
  F.~Febres Cordero, L.~Reina and D.~Wackeroth,
  %``NLO QCD corrections to Z b anti-b production with massive bottom quarks at
  %the Fermilab Tevatron,''
  Phys.\ Rev.\  D {\bf 78}, 074014 (2008)
  [arXiv:0806.0808 [hep-ph]].
  %%CITATION = PHRVA,D78,074014;%%

%\cite{Ciccolini:2007ec}
\bibitem{Ciccolini:2007ec}
  M.~Ciccolini, A.~Denner and S.~Dittmaier,
  %``Electroweak and QCD corrections to Higgs production via vector-boson fusion
  %at the LHC,''
  arXiv:0710.4749 [hep-ph].
  %%CITATION = ARXIV:0710.4749;%%

%\cite{Ciccolini:2007jr}
\bibitem{Ciccolini:2007jr}
  M.~Ciccolini, A.~Denner and S.~Dittmaier,
  %``Strong and electroweak corrections to the production of Higgs+2jets via
  %weak interactions at the LHC,''
  Phys.\ Rev.\ Lett.\  {\bf 99} (2007) 161803
  [arXiv:0707.0381 [hep-ph]].
  %%CITATION = PRLTA,99,161803;%%

%\cite{Weber:2006au}
\bibitem{Weber:2006au}
  M.~M.~Weber,
  %``Gluon initiated vector boson fusion,''
  Nucl.\ Phys.\ Proc.\ Suppl.\  {\bf 160} (2006) 200.
  %%CITATION = NUPHZ,160,200;%%

%\cite{Andersen:2006ag}
\bibitem{Andersen:2006ag}
  J.~R.~Andersen and J.~M.~Smillie,
  %``QCD and electroweak interference in Higgs production by gauge boson
  %fusion,''
  Phys.\ Rev.\  D {\bf 75}, 037301 (2007)
  [arXiv:hep-ph/0611281].
  %%CITATION = PHRVA,D75,037301;%%

%\cite{Andersen:2007mp}
\bibitem{Andersen:2007mp}
  J.~R.~Andersen, T.~Binoth, G.~Heinrich and J.~M.~Smillie,
  %``Loop induced interference effects in Higgs Boson plus two jet production at
  %the LHC,''
  arXiv:0709.3513 [hep-ph].
  %%CITATION = ARXIV:0709.3513;%%

%\cite{Bredenstein:2008tm}
\bibitem{Bredenstein:2008tm}
  A.~Bredenstein, K.~Hagiwara and B.~Jager,
  %``Mixed QCD-electroweak contributions to Higgs-plus-dijet production at the
  %LHC,''
  Phys.\ Rev.\  D {\bf 77} (2008) 073004
  [arXiv:0801.4231 [hep-ph]].
  %%CITATION = PHRVA,D77,073004;%%

%\cite{Bern:2008ef}
\bibitem{Bern:2008ef}
  Z.~Bern {\it et al.}  [NLO Multileg Working Group],
  %``The NLO multileg working group: summary report,''
  arXiv:0803.0494 [hep-ph].
  %%CITATION = ARXIV:0803.0494;%%

%\cite{Bredenstein:2008zb}
\bibitem{Bredenstein:2008zb}
  A.~Bredenstein, A.~Denner, S.~Dittmaier and S.~Pozzorini,
  %``NLO QCD corrections to top anti-top bottom anti-bottom production at the
  %LHC: 1. quark-antiquark annihilation,''
  JHEP {\bf 0808} (2008) 108
  [arXiv:0807.1248 [hep-ph]].
  %%CITATION = JHEPA,0808,108;%%

%\cite{Binoth:2008gx}
\bibitem{Binoth:2008gx}
  T.~Binoth {\it et al.},
  %``Precise predictions for LHC using a GOLEM,''
  arXiv:0807.0605 [hep-ph].
  %%CITATION = ARXIV:0807.0605;%

%\cite{Ellis:2008qc}
\bibitem{Ellis:2008qc}
  R.~K.~Ellis, W.~T.~Giele, Z.~Kunszt, K.~Melnikov and G.~Zanderighi,
  %``One-loop amplitudes for W+3 jet production in hadron collisions,''
  JHEP {\bf 0901} (2009) 012
  [arXiv:0810.2762 [hep-ph]].
  %%CITATION = JHEPA,0901,012;%%

%\cite{Berger:2009zg}
\bibitem{Berger:2009zg}
  C.~F.~Berger {\it et al.},
  %``Precise Predictions for $W$ + 3 Jet Production at Hadron Colliders,''
  arXiv:0902.2760 [hep-ph].
  %%CITATION = ARXIV:0902.2760;%%

%\cite{Reiter:2009kb}
\bibitem{Reiter:2009kb}
  T.~Reiter,
  %``Automated Evaluation of One-Loop Six-Point Processes for the LHC,''
  arXiv:0903.0947 [hep-ph].
  %%CITATION = ARXIV:0903.0947;%%


%\cite{Dixon:1996wi}
\bibitem{Dixon:1996wi}
  L.~J.~Dixon, (TASI 1995),
  %``Calculating scattering amplitudes efficiently,''
  arXiv:hep-ph/9601359.
  %%CITATION = HEP-PH/9601359;%%

%\cite{Maitre:2007jq}
\bibitem{Maitre:2007jq}
  D.~Maitre and P.~Mastrolia,
  %``S@M, a Mathematica Implementation of the Spinor-Helicity Formalism,''
  arXiv:0710.5559 [hep-ph].
  %%CITATION = ARXIV:0710.5559;%%


%\cite{Kanaki:2000ms}
\bibitem{Kanaki:2000ms}
  A.~Kanaki and C.~G.~Papadopoulos,
  %``HELAC-PHEGAS: Automatic computation of helicity amplitudes and cross
  %sections,''
  arXiv:hep-ph/0012004.
  %%CITATION = HEP-PH/0012004;%%


%\cite{Maltoni:2002mq}
\bibitem{Maltoni:2002mq}
  F.~Maltoni, K.~Paul, T.~Stelzer and S.~Willenbrock,
  %``Color-flow decomposition of QCD amplitudes,''
  Phys.\ Rev.\  D {\bf 67} (2003) 014026
  [arXiv:hep-ph/0209271].
  %%CITATION = PHRVA,D67,014026;%%


%\cite{Papadopoulos:2005ky}
\bibitem{Papadopoulos:2005ky}
  C.~G.~Papadopoulos and M.~Worek,
  %``Multi-parton Cross Sections at Hadron Colliders,''
  Eur.\ Phys.\ J.\  C {\bf 50}, 843 (2007)
  [arXiv:hep-ph/0512150].
  %%CITATION = EPHJA,C50,843;%%


%\cite{Cutkosky:1960sp}
\bibitem{Cutkosky:1960sp}
  R.~E.~Cutkosky,
  %``Singularities and discontinuities of Feynman amplitudes,''
  J.\ Math.\ Phys.\  {\bf 1} (1960) 429.
  %%CITATION = JMAPA,1,429;%%

%\cite{analytic_s_mat}
\bibitem{analytic_s_mat}
 R.~J.~Eden, P.~V.~Landshoff, D.~I.~Olive, J.~C.~Polkinghorne,
  %``The analytic S-matrix,''
 (Cambridge University Press, 1966).
  %%CITATION = JMAPA,1,429;%%

%\cite{Bern:2007dw}
\bibitem{Bern:2007dw}
  Z.~Bern, L.~J.~Dixon and D.~A.~Kosower,
  %``On-Shell Methods in Perturbative QCD,''
  Annals Phys.\  {\bf 322} (2007) 1587
  [arXiv:0704.2798 [hep-ph]].
  %%CITATION = APNYA,322,1587;%%


%\cite{Berger:2006ci}
\bibitem{Berger:2006ci}
  C.~F.~Berger, Z.~Bern, L.~J.~Dixon, D.~Forde and D.~A.~Kosower,
  %``Bootstrapping one-loop QCD amplitudes with general helicities,''
  Phys.\ Rev.\  D {\bf 74} (2006) 036009
  [arXiv:hep-ph/0604195].
  %%CITATION = PHRVA,D74,036009;%%

%\cite{Anastasiou:2006jv}
\bibitem{Anastasiou:2006jv}
  C.~Anastasiou, R.~Britto, B.~Feng, Z.~Kunszt and P.~Mastrolia,
  %``D-dimensional unitarity cut method,''
  Phys.\ Lett.\  B {\bf 645} (2007) 213
  [arXiv:hep-ph/0609191].
  %%CITATION = PHLTA,B645,213;%%


%\cite{Bern:1994cg}
\bibitem{Bern:1994cg}
  Z.~Bern, L.~J.~Dixon, D.~C.~Dunbar and D.~A.~Kosower,
  %``Fusing gauge theory tree amplitudes into loop amplitudes,''
  Nucl.\ Phys.\  B {\bf 435}, 59 (1995)
  [arXiv:hep-ph/9409265].
  %%CITATION = NUPHA,B435,59;%%

%\cite{Bern:1994zx}
\bibitem{Bern:1994zx}
  Z.~Bern, L.~J.~Dixon, D.~C.~Dunbar and D.~A.~Kosower,
  %``One-Loop n-Point Gauge Theory Amplitudes, Unitarity and Collinear Limits,''
  Nucl.\ Phys.\  B {\bf 425}, 217 (1994)
  [arXiv:hep-ph/9403226].
  %%CITATION = NUPHA,B425,217;%%

%\cite{Witten:2003nn}
\bibitem{Witten:2003nn}
  E.~Witten,
  %``Perturbative gauge theory as a string theory in twistor space,''
  Commun.\ Math.\ Phys.\  {\bf 252} (2004) 189
  [arXiv:hep-th/0312171].
  %%CITATION = CMPHA,252,189;%%


%\cite{Britto:2004nc}
\bibitem{Britto:2004nc}
  R.~Britto, F.~Cachazo and B.~Feng,
  %``Generalized unitarity and one-loop amplitudes in N = 4  super-Yang-Mills,''
  Nucl.\ Phys.\  B {\bf 725}, 275 (2005)
  [arXiv:hep-th/0412103].
  %%CITATION = NUPHA,B725,275;%%

%\cite{Brandhuber:2005jw}
\bibitem{Brandhuber:2005jw}
  A.~Brandhuber, S.~McNamara, B.~J.~Spence and G.~Travaglini,
  %``Loop amplitudes in pure Yang-Mills from generalised unitarity,''
  JHEP {\bf 0510}, 011 (2005)
  [arXiv:hep-th/0506068].
  %%CITATION = JHEPA,0510,011;%%

%\cite{Britto:2006sj}
\bibitem{Britto:2006sj}
  R.~Britto, B.~Feng and P.~Mastrolia,
  %``The cut-constructible part of QCD amplitudes,''
  Phys.\ Rev.\  D {\bf 73}, 105004 (2006)
  [arXiv:hep-ph/0602178].
  %%CITATION = PHRVA,D73,105004;%%

%\cite{Dunbar:2008zz}
\bibitem{Dunbar:2008zz}
  D.~C.~Dunbar,
  %``The six gluon one-loop amplitude,''
  Nucl.\ Phys.\ Proc.\ Suppl.\  {\bf 183}, 122 (2008).
  %%CITATION = NUPHZ,183,122;%%

%\cite{Dunbar:2009uk}
\bibitem{Dunbar:2009uk}
  D.~C.~Dunbar,
  %``One-Loop Six Gluon Amplitude,''
  arXiv:0901.1202 [hep-ph].
  %%CITATION = ARXIV:0901.1202;%%

%\cite{Xiao:2006vt}
\bibitem{Xiao:2006vt}
  Z.~Xiao, G.~Yang and C.~J.~Zhu,
  %``The rational parts of one-loop QCD amplitudes III: The six-gluon case,''
  Nucl.\ Phys.\  B {\bf 758}, 53 (2006)
  [arXiv:hep-ph/0607017].
  %%CITATION = NUPHA,B758,53;%%

%\cite{Su:2006vs}
\bibitem{Su:2006vs}
  X.~Su, Z.~Xiao, G.~Yang and C.~J.~Zhu,
  %``The rational part of QCD amplitude. II: The five-gluon,''
  Nucl.\ Phys.\  B {\bf 758}, 35 (2006)
  [arXiv:hep-ph/0607016].
  %%CITATION = NUPHA,B758,35;%%

%\cite{Xiao:2006vr}
\bibitem{Xiao:2006vr}
  Z.~Xiao, G.~Yang and C.~J.~Zhu,
  %``The rational part of QCD amplitude. I: The general formalism,''
  Nucl.\ Phys.\  B {\bf 758}, 1 (2006)
  [arXiv:hep-ph/0607015].
  %%CITATION = NUPHA,B758,1;%%

%\cite{Binoth:2007ca}
\bibitem{Binoth:2007ca}
  T.~Binoth, G.~Heinrich, T.~Gehrmann and P.~Mastrolia,
  %``Six-Photon Amplitudes,''
  Phys.\ Lett.\  B {\bf 649} (2007) 422
  [arXiv:hep-ph/0703311].
  %%CITATION = PHLTA,B649,422;%%

%\cite{Ossola:2007bb}
\bibitem{Ossola:2007bb}
  G.~Ossola, C.~G.~Papadopoulos and R.~Pittau,
  %``Numerical Evaluation of Six-Photon Amplitudes,''
  JHEP {\bf 0707} (2007) 085
  [arXiv:0704.1271 [hep-ph]].
  %%CITATION = JHEPA,0707,085;%%

%\cite{Binoth:2006hk}
\bibitem{Binoth:2006hk}
  T.~Binoth, J.~P.~Guillet and G.~Heinrich,
  %``Algebraic evaluation of rational polynomials in one-loop amplitudes,''
  JHEP {\bf 0702} (2007) 013
  [arXiv:hep-ph/0609054].
  %%CITATION = JHEPA,0702,013;%%

%\cite{Berger:2008sj}
\bibitem{Berger:2008sj}
  C.~F.~Berger {\it et al.},
  %``An Automated Implementation of On-Shell Methods for One-Loop Amplitudes,''
  Phys.\ Rev.\  D {\bf 78} (2008) 036003
  [arXiv:0803.4180 [hep-ph]].
  %%CITATION = PHRVA,D78,036003;%%

%\cite{Berger:2008ag}
\bibitem{Berger:2008ag}
  C.~F.~Berger {\it et al.},
  %``One-Loop Calculations with BlackHat,''
  Nucl.\ Phys.\ Proc.\ Suppl.\  {\bf 183} (2008) 313
  [arXiv:0807.3705 [hep-ph]].
  %%CITATION = NUPHZ,183,313;%%

%\cite{Ellis:2007br}
\bibitem{Ellis:2007br}
  R.~K.~Ellis, W.~T.~Giele and Z.~Kunszt,
  %``A Numerical Unitarity Formalism for Evaluating One-Loop Amplitudes,''
  JHEP {\bf 0803}, 003 (2008)
  [arXiv:0708.2398 [hep-ph]].
  %%CITATION = JHEPA,0803,003;%%

%\cite{Berger:2008sz}
\bibitem{Berger:2008sz}
  C.~F.~Berger {\it et al.},
  %``One-Loop Multi-Parton Amplitudes with a Vector Boson for the LHC,''
  arXiv:0808.0941 [hep-ph].
  %%CITATION = ARXIV:0808.0941;%%


%\cite{bailey_talk}
\bibitem{bailey_talk}
 D.~H.~Bailey, these proceedings.\\
http://crd.lbl.gov/~dhbailey/
  %
  %%CITATION = 

%\cite{Ossola:2006us}
\bibitem{Ossola:2006us}
  G.~Ossola, C.~G.~Papadopoulos and R.~Pittau,
  %``Reducing full one-loop amplitudes to scalar integrals at the integrand
  %level,''
  Nucl.\ Phys.\  B {\bf 763} (2007) 147
  [arXiv:hep-ph/0609007].
  %%CITATION = NUPHA,B763,147;%%


%\cite{Ossola:2008xq}
\bibitem{Ossola:2008xq}
  G.~Ossola, C.~G.~Papadopoulos and R.~Pittau,
  %``On the Rational Terms of the one-loop amplitudes,''
  JHEP {\bf 0805}, 004 (2008)
  [arXiv:0802.1876 [hep-ph]].
  %%CITATION = JHEPA,0805,004;%%

%\cite{Draggiotis:2009yb}
\bibitem{Draggiotis:2009yb}
  P.~Draggiotis, M.~V.~Garzelli, C.~G.~Papadopoulos and R.~Pittau,
  %``Feynman Rules for the Rational Part of the QCD 1-loop amplitudes,''
  arXiv:0903.0356 [hep-ph].
  %%CITATION = ARXIV:0903.0356;%%


%\cite{Ossola:2007ax}
\bibitem{Ossola:2007ax}
  G.~Ossola, C.~G.~Papadopoulos and R.~Pittau,
  %``CutTools: a program implementing the OPP reduction method to compute
  %one-loop amplitudes,''
  JHEP {\bf 0803}, 042 (2008)
  [arXiv:0711.3596 [hep-ph]].
  %%CITATION = JHEPA,0803,042;%%


%\cite{Ellis:2008kd}
\bibitem{Ellis:2008kd}
  R.~K.~Ellis, W.~T.~Giele and Z.~Kunszt,
  %``A Numerical Unitarity Formalism for One-Loop Amplitudes,''
  PoS {\bf RADCOR2007} (2007) 020
  [arXiv:0802.4227 [hep-ph]].
  %%CITATION = POSCI,RADCOR2007,020;%%

%\cite{Ellis:2008ir}
\bibitem{Ellis:2008ir}
  R.~K.~Ellis, W.~T.~Giele, Z.~Kunszt and K.~Melnikov,
  %``Masses, fermions and generalized $D$-dimensional unitarity,''
  arXiv:0806.3467 [hep-ph].
  %%CITATION = ARXIV:0806.3467;%%

%\cite{Giele:2008bc}
\bibitem{Giele:2008bc}
  W.~T.~Giele and G.~Zanderighi,
  %``On the Numerical Evaluation of One-Loop Amplitudes: The Gluonic Case,''
  arXiv:0805.2152 [hep-ph].
  %%CITATION = ARXIV:0805.2152;%%


%\cite{Lazopoulos:2008ex}
\bibitem{Lazopoulos:2008ex}
  A.~Lazopoulos,
  %``Multi-gluon one-loop amplitudes numerically,''
  arXiv:0812.2998 [hep-ph].
  %%CITATION = ARXIV:0812.2998;%%

%\cite{Winter:2009kd}
\bibitem{Winter:2009kd}
  J.~C.~Winter and W.~T.~Giele,
  %``Calculating gluon one-loop amplitudes numerically,''
  arXiv:0902.0094 [hep-ph].
  %%CITATION = ARXIV:0902.0094;%%





%\cite{Ellis:2009zw}
\bibitem{Ellis:2009zw}
  R.~K.~Ellis, K.~Melnikov and G.~Zanderighi,
  %``Generalized unitarity at work: first NLO QCD results for hadronic W+3jet
  %production,''
  arXiv:0901.4101 [hep-ph].
  %%CITATION = ARXIV:0901.4101;%%

%\cite{Catani:2008xa}
\bibitem{Catani:2008xa}
  S.~Catani, T.~Gleisberg, F.~Krauss, G.~Rodrigo and J.~C.~Winter,
  %``From loops to trees by-passing Feynman's theorem,''
  JHEP {\bf 0809} (2008) 065
  [arXiv:0804.3170 [hep-ph]].
  %%CITATION = JHEPA,0809,065;%%


%\cite{Passarino:1978jh}
\bibitem{Passarino:1978jh}
  G.~Passarino and M.~J.~G.~Veltman,
  %``One Loop Corrections For E+ E- Annihilation Into Mu+ Mu- In The Weinberg
  %Model,''
  Nucl.\ Phys.\  B {\bf 160}, 151 (1979).
  %%CITATION = NUPHA,B160,151;%%

%\cite{Denner:1991kt}
\bibitem{Denner:1991kt}
  A.~Denner,
  %``Techniques for calculation of electroweak radiative corrections at the one
  %loop level and results for W physics at LEP-200,''
  Fortsch.\ Phys.\  {\bf 41} (1993) 307
  [arXiv:0709.1075 [hep-ph]].
  %%CITATION = FPYKA,41,307;%%

%\cite{Davydychev:1991va}
\bibitem{Davydychev:1991va}
  A.~I.~Davydychev,
  %``A Simple formula for reducing Feynman diagrams to scalar integrals,''
  Phys.\ Lett.\  B {\bf 263} (1991) 107.
  %%CITATION = PHLTA,B263,107;%%


%\cite{Bern:1992em}
\bibitem{Bern:1992em}
  Z.~Bern, L.~J.~Dixon and D.~A.~Kosower,
  %``Dimensionally Regulated One-Loop Integrals,''
  Phys.\ Lett.\  B {\bf 302} (1993) 299
  [Erratum-ibid.\  B {\bf 318} (1993) 649]
  [arXiv:hep-ph/9212308].
  %%CITATION = PHLTA,B302,299;%%

%\cite{Bern:1993kr}
\bibitem{Bern:1993kr}
  Z.~Bern, L.~J.~Dixon and D.~A.~Kosower,
  %``Dimensionally regulated pentagon integrals,''
  Nucl.\ Phys.\  B {\bf 412} (1994) 751
  [arXiv:hep-ph/9306240].
  %%CITATION = NUPHA,B412,751;%%

%\cite{Tarasov:1996br}
\bibitem{Tarasov:1996br}
  O.~V.~Tarasov,
  %``Connection between Feynman integrals having different values of the
  %space-time dimension,''
  Phys.\ Rev.\  D {\bf 54} (1996) 6479
  [arXiv:hep-th/9606018].
  %%CITATION = PHRVA,D54,6479;%%

%\cite{Binoth:1999sp}
\bibitem{Binoth:1999sp}
  T.~Binoth, J.~P.~Guillet and G.~Heinrich,
  %``Reduction formalism for dimensionally regulated one-loop N-point
  %integrals,''
  Nucl.\ Phys.\  B {\bf 572}, 361 (2000)
  [arXiv:hep-ph/9911342].
  %%CITATION = NUPHA,B572,361;%%


%\cite{Duplancic:2003tv}
\bibitem{Duplancic:2003tv}
  G.~Duplancic and B.~Nizic,
  %``Reduction method for dimensionally regulated one-loop N-point Feynman
  %integrals,''
  Eur.\ Phys.\ J.\  C {\bf 35} (2004) 105
  [arXiv:hep-ph/0303184].
  %%CITATION = EPHJA,C35,105;%%

%\cite{Giele:2004iy}
\bibitem{Giele:2004iy}
  W.~T.~Giele and E.~W.~N.~Glover,
  %``A calculational formalism for one-loop integrals,''
  JHEP {\bf 0404} (2004) 029
  [arXiv:hep-ph/0402152].
  %%CITATION = JHEPA,0404,029;%%



%\cite{Ellis:2007qk}
\bibitem{Ellis:2007qk}
  R.~K.~Ellis and G.~Zanderighi,
  %``Scalar one-loop integrals for QCD,''
  JHEP {\bf 0802} (2008) 002
  [arXiv:0712.1851 [hep-ph]].
  %%CITATION = JHEPA,0802,002;%%

%\cite{vanOldenborgh:1990yc}
\bibitem{vanOldenborgh:1990yc}
  G.~J.~van Oldenborgh,
  %``FF: A Package to evaluate one loop Feynman diagrams,''
  Comput.\ Phys.\ Commun.\  {\bf 66}, 1 (1991).
  %%CITATION = CPHCB,66,1;%%

%\cite{vanHameren:2005ed}
\bibitem{vanHameren:2005ed}
  A.~van Hameren, J.~Vollinga and S.~Weinzierl,
  %``Automated computation of one-loop integrals in massless theories,''
  Eur.\ Phys.\ J.\  C {\bf 41}, 361 (2005)
  [arXiv:hep-ph/0502165].
  %%CITATION = EPHJA,C41,361;%%

%\cite{Binoth:2008uq}
\bibitem{Binoth:2008uq}
  T.~Binoth, J.~P.~Guillet, G.~Heinrich, E.~Pilon and T.~Reiter,
  %``Golem95: a numerical program to calculate one-loop tensor integrals with up
  %to six external legs,''
  arXiv:0810.0992 [hep-ph].
  %%CITATION = ARXIV:0810.0992;%%

%\cite{Hahn:1999mt}
\bibitem{Hahn:1999mt}
  T.~Hahn,
  %``Loop calculations with FeynArts, FormCalc, and LoopTools,''
  Acta Phys.\ Polon.\  B {\bf 30}, 3469 (1999)
  [arXiv:hep-ph/9910227].
  %%CITATION = APPOA,B30,3469;%%

%\cite{Denner:2005fg}
\bibitem{Denner:2005fg}
  A.~Denner, S.~Dittmaier, M.~Roth and L.~H.~Wieders,
  %``Electroweak corrections to charged-current e+ e- --> 4 fermion  processes:
  %Technical details and further results,''
  Nucl.\ Phys.\  B {\bf 724}, 247 (2005)
  [arXiv:hep-ph/0505042].
  %%CITATION = NUPHA,B724,247;%%

%\cite{Boudjema:2005rk}
\bibitem{Boudjema:2005rk}
  F.~Boudjema {\it et al.},
  %``Electroweak corrections for the study of the Higgs potential at the LC,''
{\it In the Proceedings of 2005 International Linear Collider Workshop (LCWS 2005), Stanford, California, 18-22 Mar 2005, pp 0601}
  [arXiv:hep-ph/0510184].
  %%CITATION = ECONF,C050318,0601;%%

%\cite{Lei:2007rv}
\bibitem{Lei:2007rv}
  G.~Lei, M.~Wen-Gan, H.~Liang, Z.~Ren-You and J.~Yi,
  %``QCD corrections to t\bar t b \bar b productions via photon-photon
  %collisions at linear colliders,''
  arXiv:0708.2951 [hep-ph].
  %%CITATION = ARXIV:0708.2951;%%

%\cite{Denner:2005nn}
\bibitem{Denner:2005nn}
  A.~Denner and S.~Dittmaier,
  %``Reduction schemes for one-loop tensor integrals,''
  Nucl.\ Phys.\  B {\bf 734} (2006) 62
  [arXiv:hep-ph/0509141].
  %%CITATION = NUPHA,B734,62;%%


%\cite{Binoth:2005ff}
\bibitem{Binoth:2005ff}
  T.~Binoth, J.~P.~Guillet, G.~Heinrich, E.~Pilon and C.~Schubert,
  %``An algebraic / numerical formalism for one-loop multi-leg amplitudes,''
  JHEP {\bf 0510} (2005) 015
  [arXiv:hep-ph/0504267].
  %%CITATION = JHEPA,0510,015;%%

%\cite{Diakonidis:2008ij}
\bibitem{Diakonidis:2008ij}
  T.~Diakonidis, J.~Fleischer, J.~Gluza, K.~Kajda, T.~Riemann and J.~B.~Tausk,
  %``A complete reduction of one-loop tensor 5- and 6-point integrals,''
  arXiv:0812.2134 [hep-ph].
  %%CITATION = ARXIV:0812.2134;%%

%\cite{Diakonidis:2009yu}
\bibitem{Diakonidis:2009yu}
  T.~Diakonidis,
  %``Reduction Method for One-loop Tensor 5- and 6-point Integrals Revisited,''
  arXiv:0901.4455 [hep-ph].
  %%CITATION = ARXIV:0901.4455;%%

%\cite{Nogueira:1991ex}
\bibitem{Nogueira:1991ex}
  P.~Nogueira,
  %``Automatic Feynman graph generation,''
  J.\ Comput.\ Phys.\  {\bf 105}, 279 (1993).
  %%CITATION = JCTPA,105,279;%%


%\cite{Vermaseren:2000nd}
\bibitem{Vermaseren:2000nd}
  J.~A.~M.~Vermaseren,
  %``New features of FORM,''
  arXiv:math-ph/0010025.
  %%CITATION = MATH-PH/0010025;%%

%\cite{Binoth:2003xk}
\bibitem{Binoth:2003xk}
  T.~Binoth, J.~P.~Guillet and F.~Mahmoudi,
  %``A compact representation of the gamma gamma g g g --> 0 amplitude,''
  JHEP {\bf 0402}, 057 (2004)
  [arXiv:hep-ph/0312334].
  %%CITATION = JHEPA,0402,057;%%

%\cite{Nagy:2006xy}
\bibitem{Nagy:2006xy}
  Z.~Nagy and D.~E.~Soper,
  %``Numerical integration of one-loop Feynman diagrams for N-photon
  %amplitudes,''
  Phys.\ Rev.\  D {\bf 74} (2006) 093006
  [arXiv:hep-ph/0610028].
  %%CITATION = PHRVA,D74,093006;%%


%\cite{Bernicot:2007hs}
\bibitem{Bernicot:2007hs}
  C.~Bernicot and J.~P.~Guillet,
  %``Six-Photon Amplitudes in Scalar QED,''
  JHEP {\bf 0801} (2008) 059
  [arXiv:0711.4713 [hep-ph]].
  %%CITATION = JHEPA,0801,059;%%

%\cite{Gong:2008ww}
\bibitem{Gong:2008ww}
  W.~Gong, Z.~Nagy and D.~E.~Soper,
  %``Direct numerical integration of one-loop Feynman diagrams for N-photon
  %amplitudes,''
  Phys.\ Rev.\  D {\bf 79}, 033005 (2009)
  [arXiv:0812.3686 [hep-ph]].
  %%CITATION = PHRVA,D79,033005;%%

%\cite{Ferroglia:2002mz}
\bibitem{Ferroglia:2002mz}
  A.~Ferroglia, M.~Passera, G.~Passarino and S.~Uccirati,
  %``All-purpose numerical evaluation of one-loop multi-leg Feynman diagrams,''
  Nucl.\ Phys.\  B {\bf 650} (2003) 162
  [arXiv:hep-ph/0209219].
  %%CITATION = NUPHA,B650,162;%%
 
  %\cite{Binoth:2002xh}
\bibitem{Binoth:2002xh}
  T.~Binoth, G.~Heinrich and N.~Kauer,
  %``A numerical evaluation of the scalar hexagon integral in the physical
  %region,''
  Nucl.\ Phys.\  B {\bf 654} (2003) 277
  [arXiv:hep-ph/0210023].
  %%CITATION = NUPHA,B654,277;%%

%\cite{Kurihara:2005ja}
\bibitem{Kurihara:2005ja}
  Y.~Kurihara and T.~Kaneko,
  %``Numerical contour integration for loop integrals,''
  Comput.\ Phys.\ Commun.\  {\bf 174} (2006) 530
  [arXiv:hep-ph/0503003].
  %%CITATION = CPHCB,174,530;%%

%\cite{Anastasiou:2007qb}
\bibitem{Anastasiou:2007qb}
  C.~Anastasiou, S.~Beerli and A.~Daleo,
  %``Evaluating multi-loop Feynman diagrams with infrared and threshold
  %singularities numerically,''
  JHEP {\bf 0705} (2007) 071
  [arXiv:hep-ph/0703282].
  %%CITATION = JHEPA,0705,071;%%



\end{thebibliography}
\end{document}